\documentclass[aps, prb, twocolumn, amsmath, amssymb, nofootinbib]{revtex4-2}
\usepackage[colorlinks]{hyperref}
\hypersetup{hidelinks}
\usepackage{bm}
\usepackage{xcolor}
\usepackage{amsmath}
\usepackage{amsfonts}
\usepackage{dsfont}
\usepackage{graphics}
\usepackage[caption=false]{subfig}
\usepackage{overpic}
\usepackage[capitalise]{cleveref}
\usepackage{tikz}
\usepackage{simpler-wick}
\usepackage{physics}
\usepackage{braket}
\usepackage{microtype}

\makeatletter
\renewcommand{\maketag@@@}[1]{\hbox{\m@th\normalsize\normalfont#1}}%
\makeatother

\newcommand{\mean}[1]{\left\lanFright\rangle}

\newcommand{\T}{\mathrm{T}}

\begin{document}
\title{Theory of Anderson transition in three-dimensional chiral symmetry classes: \\ Connection to type-II superconductors}
\author{Pengwei Zhao}
\affiliation{International Center for Quantum Materials, Peking University, Beijing 100871, China}
\author{Ryuichi Shindou}
\email{rshindou@pku.edu.cn}
\affiliation{International Center for Quantum Materials, Peking University, Beijing 100871, China}
\date{\today}

\begin{abstract}
Phase transitions governed by topological defects constitute a cornerstone of modern physics. Two-dimensional (2D) Anderson transitions in chiral symmetry classes are driven by the proliferation of vortex-antivortex pairs—a mechanism analogous to the Berezinskii–Kosterlitz–Thouless (BKT) transition in the 2D XY model. In this work, we extend this paradigm to three-dimensional (3D) chiral symmetry classes, where vortex loops emerge as the key topological defects governing the Anderson transition. By deriving the dual representation of the 3D nonlinear sigma model for the chiral unitary class, we develop a mean-field theory of its Anderson transition and elucidate the role of one-dimensional (1D) weak band topology in the Anderson transition. Strikingly, our dual representation of the 3D NLSM in the chiral symmetry class uncovers its connection to the magnetostatics of 3D type-II superconductors. The metal-to-quasi-localized and quasi-localized-to-insulating transitions in 3D chiral symmetry class share a unified theoretical framework with the normal-to-mixed and mixed-to-superconducting transitions in 3D type-II superconductors under an external magnetic field, respectively. 
\end{abstract}

\maketitle

\section{Introduction}
Topological defects are universal entities in physical systems with symmetries~\cite{mermin1979,chaikin_lubensky1995}. Characterized by quantized charges, topological defects exhibit remarkable stability against local perturbations. Well-known examples include domain walls in Ising magnets~\cite{kittel1946,sethna2006}, vortices in superconductors and superfluids~\cite{onsager1949,feynman1955,abrikosov1957,tinkham1996,donnelly1991}, and skyrmions in Heisenberg magnets~\cite{skyrme1961,skyrme1962,bogdanov1989,nagaosa2013,fert2017}. Thermodynamic properties of physical systems are profoundly influenced by topological defects, whose spatial proliferation introduces disorder in an ordered phase, leading to dramatic changes in the macroscopic properties of the systems. This mechanism drives fundamental phase transitions, such as Berezinskii–Kosterlitz–Thouless (BKT) transition~\cite{berezinskii71,berezinskii72,kosterlitz1973,kosterlitz1974}, melting transitions of 2D solids~\cite{nelson1979,young1979,halperin1978}, and the superfluid transition~\cite{onsager1949,feynman1955,nelson1977,dasgupta1981,williams1987,shenoy1989}, underscoring the pivotal role of topological defects in shaping collective phenomena~\cite{kibble1976,zurek1985,zurek1996}. 

In disordered electron systems---such as an electron gas with random impurities---charge carriers experience repeated scattering events that reduce their mean-free time. In the weak disorder regime, electrons retain sufficient coherence to propagate diffusively, resulting in a metallic phase with finite conductivity. When disorder strength exceeds a critical disorder strength, the random potentials cause spatial localization of electronic states~\cite{anderson1958,thouless1974,abrahams1979}. This metal-insulator transition, known as the Anderson transition, has been experimentally studied in diverse wave systems, which include not only electron waves in solid-state materials~\cite{rosenbaum1980,tsui1982} but also electromagnetic and atomic waves in optical, photonic, and cold-atom systems~\cite{wang2011,hu2008,roati2008,skipetrov2016}.     

Decades of theoretical and numerical studies~\cite{wegner1976,wegner1979,efetov1980,hikami1980,hikami1981,efetov1983,mackinnon1981,mackinnon1983,altlandNonstandardSymmetryClasses1997,slevin1999,eversAndersonTransitions2008,slevinCriticalExponentAnderson2014}  have demonstrated that the Anderson transitions exhibit universal scaling behaviors characterized by critical exponents, which define distinct universality classes. The universality classes are classified according to fundamental Hamiltonian symmetries---time-reversal (TRS), particle-hole (PHS), and chiral (CS) symmetries---which constitute the celebrated Altland-Zirnbauer ten-fold way~\cite{altlandNonstandardSymmetryClasses1997}. The classification scheme has since been extended to non-Hermitian matrices~\cite{kawabataSymmetryTopologyNonHermitian2019a,bernard2002,zhou2019} and their eigenstates' localization phenomena~\cite{kawabata2021,luoUnifyingAndersonTransitions2022,chen2025}, revealing its broad applicability in the study of non-Hermitian disordered systems~\cite{hatano1996,cao1999,wersma2008,wersma2013,yamilov2014,basiri2014,xu2016,yusipov2017,hamazaki2019,tzortzakakis2020,wang2020,huang2020a,huang2020b,weidemann2021,claes2021,liuzhoushu2021,liu2021,luo2021a,luoTransferMatrixStudy2021,luoUnifyingAndersonTransitions2022,tomasi2023,wang2023,huang2025,mo2025}. 

Despite decades of research, the nature of the Anderson transition in some symmetry classes remains enigmatic~\cite{gadeReplicaLimitUnSOn1991,gadeAndersonLocalizationSublattice1993,balentsDelocalizationTransitionSupersymmetry1997,hatsugai1997,altland1999,fukui1999,fabrizio2000,guruswamy2000,motrunich2002,horovitz2002,mudry2003,bocquet2003,yamada2004,garcia-garcia2006,dellanna2006,markos2007,konigMetalinsulatorTransitionTwodimensional2012,mondragon-shem2014,luoCriticalBehaviorAnderson2020,li2020,wang2021,karcherMetalinsulatorTransitionTwodimensional2023,karcher2023b,xiaoAnisotropicTopologicalAnderson2023,zhaoTopologicalEffectAnderson2024c,silva2025}. The Anderson transition has been studied by using effective field theories for disordered quantum systems---nonlinear sigma models (NLSMs)---in which the coupling constant corresponds to conductivity, and the $\beta$ function of the conductivity determines the presence (or absence) of the Anderson transition as well as its universality class~\cite{wegner1976,wegner1979,efetov1980,hikami1980,hikami1981,efetov1983,eversAndersonTransitions2008}. In the early 90s, Gade and Wegner found that the perturbative $\beta$ function for chiral symmetry classes (AIII, CII, and BDI) has no localization effects in the dimension $D\ge 2$, implying that metallic phases in these classes remain stable regardless of disorder strength—contrary to the expectation that sufficiently strong disorder should always induce the Anderson transition~\cite{gadeAndersonLocalizationSublattice1993, gadeReplicaLimitUnSOn1991}. The puzzle was later resolved in 2D by K\"onig et al.~\cite{konigMetalinsulatorTransitionTwodimensional2012}, who incorporated non-perturbative effects arising from topological defects in the NLSMs. The NLSM field variable in chiral symmetry classes resides in curved manifolds of the unitary group. Such manifolds inherently possess a U(1) subgroup symmetry, permitting vortex solutions. Thus, the 2D Anderson transition in chiral symmetry classes can be understood as a quantum phase transition driven by the spatial proliferation of U(1) vortex-antivortex pairs~\cite{konigMetalinsulatorTransitionTwodimensional2012,zhaoTopologicalEffectAnderson2024c}. In contrast, the theoretical framework of the 3D Anderson transition in chiral symmetry classes remains largely unexplored, despite numerical confirmations of its existence~\cite{garcia-garcia2006,luoCriticalBehaviorAnderson2020,wang2021,xiaoAnisotropicTopologicalAnderson2023}. 

In this paper, we generalize the 2D theory~\cite{konigMetalinsulatorTransitionTwodimensional2012} into three-dimensional (3D) chiral symmetry classes, where the NLSM field variable permits vortex {\it loop} solutions. We argue that the 3D Anderson transition in chiral classes is driven by the spatial proliferation of the vortex loops. We derive a dual representation of the 3D NLSM for the chiral unitary class and construct a mean-field theory of its 3D Anderson transition. The dual model takes the form of $\mathrm{U}(N)$ type-II superconductors.  
\begin{itemize}
\item The diffusive metal phase with a sparse population of vortex loops is described as the normal (Maxwell) phase in the type-II superconductors, where magnetic gauge field fluctuations---driven by the entropic effect but controlled by the Maxwell action---suppress superconductivity entirely.
\item The localized phase, signaled by vortex-loop proliferation, is described as the superconducting (Meissner) phase, where the Josephson coupling stabilizes superconducting order, and the spontaneous symmetry breaking of the global U(1) gauge symmetry renders the magnetic gauge field massive via the Anderson-Higgs mechanism~\cite{anderson1963,higgs1964}.
\end{itemize}

Recent studies have demonstrated that nontrivial 1D weak band topology in chiral symmetry classes~\cite{Ryu2010,schnyder2011,fulga2012a} gives rise to an intermediate quasi-localized phase, bridging metallic and fully localized phases~\cite{xiaoAnisotropicTopologicalAnderson2023, zhaoTopologicalEffectAnderson2024c}. In the quasi-localized phase, the exponential localization length diverges only along the weak topological direction while remaining finite in the other directions. We show that, in the dual representation, the 1D weak topology manifests as an external magnetic field applied to the type-II superconductors along the topological direction. Thereby, 
\begin{itemize}
\item the quasi-localized phase---characterized by divergent localization length only along the topological direction---can be described as an intermediate mixed phase in which the external magnetic field penetrates through the 3D superconducting bulk in the form of magnetic flux tubes~\cite{blatter1994,brandt1995,zeldov1995}.
\end{itemize}

The remainder of this paper is organized as follows. Section \ref{sec:NLSM} introduces the NLSM for chiral symmetry classes. In Appendix \ref{app:derive_NLSM}, we outline a derivation of the NLSM for the chiral unitary class.  Section \ref{sec:defects} and Appendix \ref{app:saddle_point} explain the topological defects in the 3D NLSM in the chiral symmetry classes and their qualitative impact on the localization properties of chiral symmetry classes. Section \ref{sec:duality} develops the dual representation of the NLSM. Section \ref{sec:anderson} and Appendix \ref{app:variational_method} analyze the dual lattice model without the 1D weak topology. Section \ref{sec:quasilocalization} analyzes the dual lattice model with the 1D weak topology and discusses the nature of the quasi-localized phase. Section \ref{sec:discussion} summarizes the paper. 

\section{Nonlinear sigma model}\label{sec:NLSM}
A disordered system considered here consists of a free-fermion Hamiltonian perturbed by a single-particle random potential, where the random potential follows a certain probability distribution. Physical observables in such systems are determined through two distinct averaging procedures. One is the quantum-mechanical average $\partial \ln \mathcal{Z}/\partial J$, where $\mathcal{Z}$ is a partition function of the free-particle Hamiltonian with some potential, and $J$ is a current operator that couples linearly with the observables. The other is a subsequent disorder average, $\langle \ln \mathcal{Z} \rangle_{\rm dis}$, over the potential's probability distribution. To evaluate $\langle \ln \mathcal{Z}\rangle_{\rm dis}$, three formalisms are commonly employed: replica formulation (used here)~\cite{wegner1976,wegner1979,efetov1980,altlandCondensedMatterField2010}, supersymmetry technique~\cite{efetov1983,efetovSupersymmetryDisorderChaos1999}, and Keldysh path integral formalism~\cite{kamenevManybodyTheoryNonequilibrium2005}.
The replica formulation exploits the mathematical identity:
\begin{equation}
    \ln \mathcal{Z}=\lim_{N\to 0}\frac{\mathcal{Z}^N-1}{N}. \label{eq:replica}
\end{equation}
For a Gaussian-distributed random potential, the disorder average of the replicated partition function $\langle \mathcal{Z}^N\rangle_{\rm dis}$ can be computed exactly. Upon integrating the disorder potential over the Gaussian distribution function, the system acquires an effective interaction among $N$ replicated free-fermion fields. This interaction is then decoupled by the use of a Hubbard-Stratonovich matrix field ($Q$ field), leading to a self-consistent Born equation for the fermion mean free time $\tau$.

In the weak disorder regime, the fermion has a finite mean-free time $\tau$, which spontaneously breaks continuous symmetries in a space of the $N$ replicated fields~\cite{wegner1976,wegner1979,efetov1980,altland1999}. Low-energy thermodynamics in the weak disorder regime is governed by gapless Goldstone (diffusion) modes arising from the continuous symmetry breaking, and it can be studied in terms of the spatial gradient expansion of the slowly-varying $Q$-field. Up to the second-order gradient expansion, an effective action for the Goldstone modes in chiral symmetry classes is described by the following NLSM~\cite{gadeAndersonLocalizationSublattice1993, gadeReplicaLimitUnSOn1991,fukui1999,altland1999,fabrizio2000, konigMetalinsulatorTransitionTwodimensional2012, zhaoTopologicalEffectAnderson2024c}: 
\begin{equation}\label{eq:NLSM}
\begin{aligned}
    S&=-\int\frac{\mathrm{d}^3r}{8\pi s}\,\sum_{\mu=x,y,z} \bigg[\sigma_{\mu} \Tr \left(Q^{-1}\nabla_{\mu} Q\right)^2\\
    &+c_{\mu}\Tr ^2(Q^{-1}\nabla_{\mu} Q)- \chi_{\mu}\Tr \left(Q^{-1}\nabla_{\mu} Q\right)\bigg],
\end{aligned}
\end{equation}
with the disorder-averaged replicated partition function $Z\equiv\langle \mathcal{Z}^N\rangle_{\text{dis}}=\int \mathrm{D}[Q]\,e^{-S}$ (see also Appendix \ref{app:derive_NLSM}). The $Q$-field resides on a curved manifold dubbed the Goldstone manifold, and $\mathrm{D}[Q]$ is a Haar measure on the curved manifold \cite{fuchsSymmetriesLieAlgebras2003}. 
For chiral unitary, symplectic, and orthogonal classes, the $Q$ field is $\mathrm{U}(N)$, $\mathrm{U}(N)/\mathrm{O}(N)$, and $\mathrm{U}(N)/\mathrm{Sp}(N)$, respectively \cite{eversAndersonTransitions2008, konigMetalinsulatorTransitionTwodimensional2012}
 (\cref{tab:NLSM_of_chiral_symmetry_classes}). 

In Eq.~(\ref{eq:NLSM}), the first term with conductivity $\sigma_{\mu}$ governs the dynamics of the gapless Goldstone modes---diffusion phenomena. The second term with the so-called Gade constant $c_{\mu}$ arises from couplings between the gapless Goldstone modes and high-energy massive modes.~\cite{altland1999,fukui1999,fabrizio2000,zhaoTopologicalEffectAnderson2024c} A real-valued vector $\bm{\chi}$ represents the 1D weak band topology, which encodes the 1D topological index of an underlying 3D lattice model~\cite{altlandQuantumCriticalityQuasiOneDimensional2014, altlandTopologyAndersonLocalization2015, zhaoTopologicalEffectAnderson2024c} (see also Appendix A3). In this paper, we choose ${\bm \chi}$ to be along the $z$ direction and refer to this direction as a topological direction. For simplicity, in the following section, we assume spatially isotropic conductivity and Gade constant: $\sigma_{\mu}=\sigma$, $c_{\mu}=c$.

\begin{table}[t]
    \centering
    \begin{tabular}{cccc}
    \hline\hline
    Symmetry class & Goldstone manifold & Extra constraint& $s$ \\
    \hline
    AIII & $\mathrm{U}(N)$ & None & 1 \\
    CII & $\mathrm{U}(N)/\mathrm{O}(N)$ & $Q=Q^\T$ & $2$ \\
    BDI & $\mathrm{U}(N)/\mathrm{Sp}(N)$ & $Q=CQ^\T C$ & $2$ \\
    \hline\hline
    \end{tabular}
    \caption{The Goldstone manifold where the field variable $Q$ lives for three chiral symmetry classes. $\mathrm{U}(N)$, $\mathrm{O}(N)$, and $\mathrm{Sp}(N)$ are the unitary, orthogonal, and symplectic groups, respectively. $C$ is a skew-symmetric matrix that satisfies $C^2=1$.}
    \label{tab:NLSM_of_chiral_symmetry_classes}
\end{table}

\section{Topological defect}\label{sec:defects}
\subsection{Saddle-point equation}
The Goldstone manifolds in all three chiral symmetry classes have a $\mathrm{U}(1)$ subgroup. An eigenvalue of the $Q$ field in such Goldstone manifolds has a vortex excitation as a stable topological defect, which takes the form of a line in 3D space~\cite{onsager1949,feynman1955}. To describe general forms of vortex excitations in the $Q$ field, consider an eigenvalue $e^{i\phi_n(\bm{r})}$ and eigenvector $\ket{p_n(\bm{r})}$ of the $Q$ field $(n=1,\cdots,N)$, respectively,
\begin{equation}
    Q(\bm{r})=\sum_{n=1}^{N}e^{i\phi_n(\bm{r})}\ket{p_n(\bm{r})}\bra{p_n(\bm{r})}.
\end{equation}
$\ket{p_n(\bm{r})}$ in the AIII class has no symmetry constraint, belonging to the complex projective space $\mathbb{CP}^{N-1}$.  The eigenvectors in the BDI class form Kramers doublet, $\ket{p_{n,\sigma}(\bm{r})}$  ($\sigma=\pm$), for each eigenvalue $e^{i\phi_n(\bm{r})}$ ($n=1,\cdots,N/2$), belonging to the quaternionic projective space $\mathbb{HP}^{N/2-1}$. The eigenvectors in the CII class are real-valued, belonging to the real projective space $\mathbb{RP}^{N-1}$\cite{konigMetalinsulatorTransitionTwodimensional2012}. Being a unit complex number, each eigenvalue $e^{i\phi_n(\bm{r})}$ permits a vortex excitation. In 3D space, a vortex excitation forms a closed line $C$ of quantized vortex -- vortex loop --, 
\begin{equation}\label{eq:winding}
\sum_{\nu,\rho}\epsilon_{\mu\nu\rho}\nabla_\nu\nabla_\rho\phi_n(\bm{r})=2\pi\eta\oint_C \mathrm{d} l_\mu\,\delta^3(\bm{r}-\bm{l}).
\end{equation}
Here $\eta \in \mathbb{Z}$ stands for vorticity, and a line integral on the right-hand side is along the vortex loop $C$. Each eigenvalue has an arbitrary number of vortex loops \cite{zhaoTopologicalEffectAnderson2024c}.

The $Q$ field with the vortex excitations can be introduced as a saddle-point solution of the chiral NLSM, Eq.~(\ref{eq:NLSM}). To this end, consider the following \emph{ansatz} as a saddle-point solution~\cite{konigMetalinsulatorTransitionTwodimensional2012,zhaoTopologicalEffectAnderson2024c}, which assumes that all the eigenvalues except for $e^{i\phi(\bm{r})}$ are degenerate, and only $e^{i\phi(\bm{r})}$ has multiple vortex loops:
\begin{equation}\label{eq:Ansatz}
    Q(\bm{r})=1+\left[e^{i\phi(\bm{r})}-1\right]\ket{p(\bm{r})}\bra{p(\bm{r})}.
\end{equation}
Here $P(\bm{r}) \equiv \ket{p(\bm{r})}\bra{p(\bm{r})}$ is a projection operator onto the one-dimensional and two-dimensional eigenspaces of $e^{i\phi(\bm{r})}$ in the AIII and CII classes, and the BDI class, respectively. A substitution of the ansatz into the NLSM action \cref{eq:NLSM} yields,
\begin{align}
    S[\phi, P]=&\frac{u}{8\pi s}\int\mathrm{d}^3r\,\sum_{\mu=x,y,z}\big[(\sigma+uc)(\nabla_{\mu}\phi)^2 \nonumber \\
    & + 2\sigma (1-\cos\phi)\Tr (\nabla_{\mu}P)^2 + i\chi_{\mu}\nabla_{\mu}\phi \big], \label{eq:action_with_Ansatz} 
\end{align}
where $u\equiv\Tr\ket{p(\bm{r})}\bra{p(\bm{r})}$ is $1$ for class AIII and CII, and $2$ for class BDI \cite{konigMetalinsulatorTransitionTwodimensional2012}. From $\delta S/\delta\phi(\bm{r})=0$ and $\delta S/\delta P(\bm{r})=0$, we obtain saddle-point equations for $\phi(\bm{r})$ and $P(\bm{r})$, respectively, 
\begin{align}
&\sum_{\mu}(\sigma+u c)\nabla^2_{\mu}\phi + \sum_{\mu}\sin\phi\, \Tr (\nabla_{\mu}P)^2=0, \label{eq:saddle-point_equation1} \\
& \sum_{\mu} \sigma (1-\cos\phi)\,\nabla^2_{\mu}P  + \sum_{\mu} \sigma \sin\phi \,\ 
\nabla_{\mu}\phi
 \nabla_{\mu} P = 0. \label{eq:saddle-point-equation2}
\end{align}
When $\phi(\bm{r})$ is definite, such as outside the vortex cores, the second equation requires that $P(\bm{r})$ is independent of $\bm{r}$ (Appendix \ref{app:saddle_point}). When $\phi(\bm{r})$ is indefinite such as at the vortex cores, $P(\bm{r})$ can depend on ${\bm r}$. Thus, the first equation yields the Poisson equation $\nabla^2\phi=0$. The Poisson equation supports saddle-point solutions with multiple vortex loops (see Fig.~\ref{fig:loops}),
\begin{align}
\sum_{\nu,\rho}\epsilon_{\mu\nu\rho}\nabla_\nu\nabla_\rho\phi(\bm{r})=2\pi
\sum^{n}_{j=1} \oint_{C_j} \mathrm{d} l_\mu\,\delta^3(\bm{r}-\bm{l}), \label{eq:winding2}
\end{align}
where $C_j$ defines the 3D coordinate of the $j$th vortex loop with the unit vorticity ($\eta_j=1$). For simplicity, we henceforth consider only vortex loops with the unit vorticity, because a vortex line with higher vorticities can be regarded as a sum of several vortex lines with the unit vorticity. As $P(\bm{r})$, except at the vortex cores, is independent of $\bm{r}$, the energy of the saddle-point solution may as well be expressed by a functional only of $\phi(\bm{r})$;
\begin{align}
&S[\phi, P]=S[\phi] \nonumber \\
&\equiv 
\frac{u}{8\pi s}\int\mathrm{d}^3r\,\sum_{\mu}\left[ (\sigma + u c) \!\ \!\ (\nabla_{\mu}\phi)^2 + i\chi_{\mu}\nabla_{\mu}\phi\right]. \label{Eq:1}
\end{align}
We note in passing that the $\bm{r}$-dependence of $P(\bm{r})$ along the vortex core may bring additional positive energies from the second term of Eq.~(\ref{eq:action_with_Ansatz}). For simplicity, in this section, we ignore such $P(\bm{r})$-dependent energy terms [see also Section \ref{sec:duality} for an alternative way of including the effect of $P(\bm{r})$].

\subsection{Vortex-driven \texorpdfstring{$Q$}{}-field fluctuation and effect of 1D weak topology}
The saddle-point equations and their solutions have a physical analogy to 3D magnetostatics in the presence of quantized electric current loops~\cite{jackson1999}. Regarding $\phi$ as a magnetic scalar potential and defining $\mathsf{h} \equiv -\bm{\nabla}\phi$, the Poisson equation becomes the magnetic Gauss law, $\bm{\nabla}\cdot \mathsf{h}=0$. Meanwhile, Eq.~(\ref{eq:winding2}) can be recast as the Biot-Savart law for $\mathsf{h}$, where a vortex loop becomes a quantized electric current loop.
    
\begin{figure}[htb]
    \centering
    \includegraphics[width=1.0\linewidth]{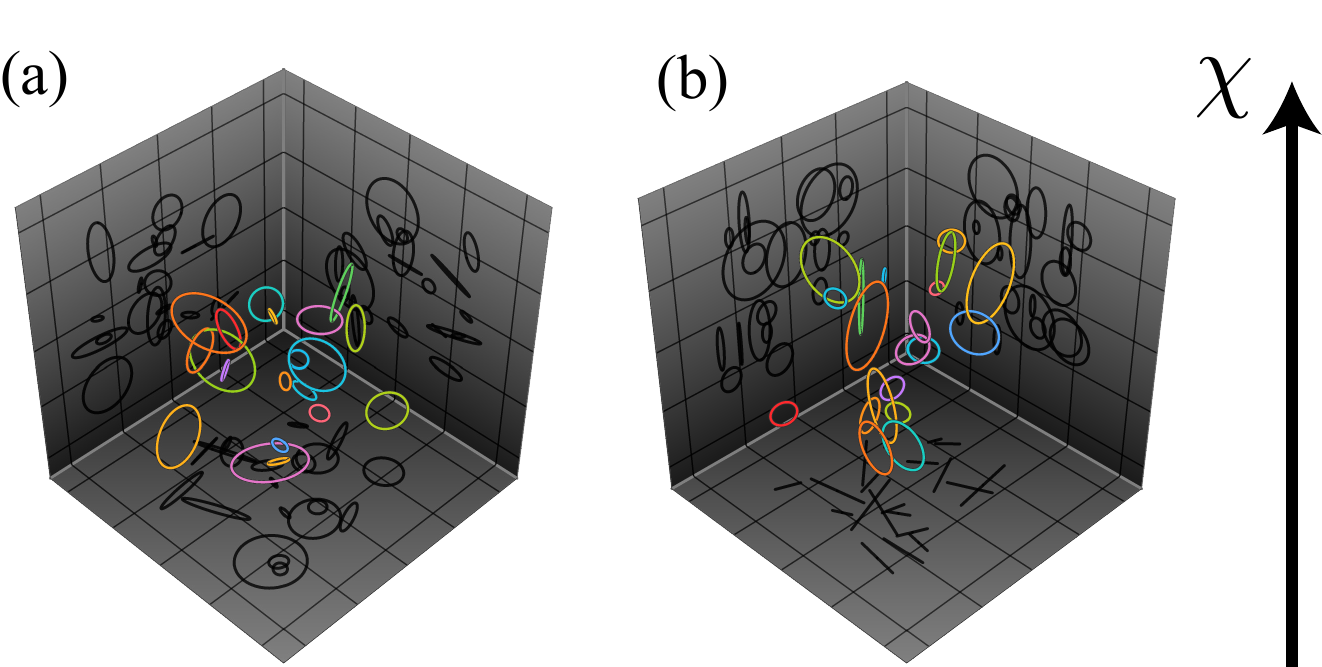}
    \caption{Schematic pictures of vortex loops in 3D. (a) Unpolarized vortex loops that have finite projections to a plane perpendicular to $\bm{\chi}$, and (b) polarized vortex loops that are confined within a plane parallel to $\bm{\chi}$.  Black-colored loops on three 2D planes are projections of the vortex loops onto the three coordinate planes. The projections of the polarized vortex loops onto the plane perpendicular to $\bm{\chi}$ enclose zero area. The figures show only circular loops, while vortex loops can take arbitrary shapes.  In the presence of the weak topological term $\bm{\chi}$, unpolarized vortex loops are suppressed by the destructive interference effect, while polarized vortex loops are free from the interference effect, dominating the partition function near the mobility edge.}
    \label{fig:loops}
\end{figure}

The analogy to the 3D magnetostatics is useful for discussing the $Q$-field fluctuation induced by the vortex excitations. The effect of vortex excitations on the $Q$-field fluctuation may be characterized by a correlation function of the U(1) phase variable, $e^{i\phi(\bm{r})}$,  
\begin{align}
F(\bm{r}-\bm{r}')\equiv \left\langle e^{i[\phi(\bm{r}')-\phi(\bm{r})]} \right\rangle. \label{Eq:2}
\end{align}
Here the average in the right-hand side is over $\phi(\bm{r})$ with and without vortex excitations, $\langle \cdots \rangle \equiv \frac{1}{Z} \int\mathrm{D}[\phi] \cdots \exp(-S[\phi])$, and $Z \equiv\int\mathrm{D}[\phi] \exp(-S[\phi])$. The fluctuation of the $\phi$ field comprises the spin-wave part $\phi_{\rm sw}$ and the vortex-excitation part $\phi_{\rm v}$. When the spin-wave part is neglected, the phase difference is rewritten by an integral of the magnetic field $\mathsf{h}$ along an arbitrary line connecting $\bm{r}$ and $\bm{r}^{\prime}$, 
\begin{align}
F(\bm{r}-\bm{r}^{\prime}) &\simeq \left\langle e^{i\int^{\bm{r}^{\prime}}_{\bm{r}}\mathrm{d}\bm{s}\cdot\mathsf{h}(\bm{s})}\right\rangle_{\rm v} \nonumber \\
&\equiv \frac{1}{Z_{\rm v}}\int \mathrm{D}[\phi_{\rm v}] \!\ e^{i\int^{\bm{r}^{\prime}}_{\bm{r}}\mathrm{d}\bm{s}\cdot\mathsf{h}(\bm{s})} e^{-S_{\rm v}}, \label{Eq:3}
\end{align}
with $Z_{\rm v} \equiv\int\mathrm{D}[\phi_{\rm v}] \!\ e^{-S_{\rm v} }$. Here the action in Eq.~(\ref{Eq:1}) is decomposed into the spin wave part $S_{\rm sw}=\int\mathrm{d}^3r/(8\pi) (\sigma+Nc) (\nabla_{\mu} \phi_{\rm sw})^2$ and vortex excitation part, $ S_{\rm v}$ with $S=S_{\rm sw}+S_{\rm v}$. From the analogy to the 3D magnetostatics, the vortex part $S_{\rm v}$ of Eq.~(\ref{Eq:1}) takes the form of a Coulomb loop gas model, where the vortex loops (electric current loops) interact with each other via the $1/r$ Coulomb interaction~\cite{popov1973,wiegel1973,savit1978,peskin1978,fradkin1978,shindou2025}. The 1D Berry phase term confers the complex phase factor upon each closed vortex loop~\cite{tanakaShortGuideTopological2015,zhaoTopologicalEffectAnderson2024c,shindou2025}, 
\begin{align}
S_{\rm v} =  g \!\ \sum_{j,m} 
\oint_{C_j} \oint_{C_m} \frac{\mathrm{d} \bm{r}_j\cdot \mathrm{d} \bm{r}_m}{|\bm{r}_j-\bm{r}_m|} 
+ i \overline{\chi}_{\mu} \sum_{j} \epsilon_{\mu\nu\rho}\Gamma_{j,\nu\rho}. \label{Eq:4}
\end{align}
Here  $g \equiv (\sigma+c)/(8s)$ and $\overline{\bm{\chi}} \equiv \bm{\chi}/(4s)$, $j,m=1,\cdots,n$, $\mu,\nu,\rho=x,y,z$, and $\mathrm{d}\bm{r}_j$ and $\mathrm{d}\bm{r}_m$ are tangential vectors of the $j$th and $m$th vortex loops $C_j$ and $C_m$ at $\bm{r}_j$ and $\bm{r}_m$, respectively. Note that the vortex loop segments (electric-current loop segments) interact via the $1/r$ Coulomb interaction. $\Gamma_{j,\nu\rho}$ is a projected area of the $j$th vortex loop onto the $\nu$-$\rho$ plane (see Fig.~\ref{fig:loops}). The path integral $\int\mathrm{D}[\phi_{\rm v}]$ over the vortex excitation part is nothing but a sum over all possible configurations of vortex loops,
\begin{align}
\int\mathrm{D} [\phi_{\rm v}]  \equiv 1 + \sum^{\infty}_{n=1} \frac{1}{n!} \prod^n_{j=1} \left(\int \mathrm{d}l_j  \!\ t^{l_j} \!\  \int \mathrm{d}^3R_j  \int \mathrm{D}\Omega_j(\lambda)\right).  \label{Eq:5}
\end{align}
The right-hand side comprises a sum over vortex-loop number $n$, an integral over a length $l_j$ of the $j$th vortex loop $C_j$ $(j=1,\cdots,n)$, an integral over a center-of-mass coordinate $R_j$ of the $j$th loop, and a path integral over the tangential vector $\Omega_j(\lambda) \equiv \mathrm{d}r_j(\lambda)/\mathrm{d}\lambda$ in the loop $C_j$ (Here $\lambda$ is a 1D length-scale parameter that parametrizes the loop segment in the loop). With a closed-loop condition of $\int^{l_j}_{0} \mathrm{d}\lambda \! \ \Omega_j(\lambda)=0$, the path integral over the tangential vector is equivalent to a summation over possible shapes of the closed loop with a fixed length $l_j$~\cite{shindou2025}. Note that a fugacity parameter $t$ of the vortex loop is newly included on the right-hand side: $\ln t$ is a chemical potential of the vortex loop segment per unit length. $1/n!$ is a symmetry factor that sets off double counting of the same multiple-vortex-loops configurations. 

The Coulomb loop gas model without the 1D Berry phase term ($\chi=0$) undergoes an order-disorder transition driven by the spatial proliferation of vortex loops (electric-current loops). The transition was previously studied by a renormalization group method, which copes with the screening effect of smaller vortex loops upon the Coulomb interaction among larger vortex loops~\cite{williams1987,shenoy1989,shindou2025}. The screening effect renormalizes the coupling constant $g$, while a short-range part of the Coulomb interaction renormalizes the fugacity parameter $\ln t$. The RG study shows that the ordered phase is characterized by a fixed point with divergent $g$ and vanishing $t$, and the disordered phase is characterized by a fixed point with vanishing $g$ and divergent $t$. In fact, the divergent $t$ results in the spatial proliferation of vortex loops (electric-current loops), which makes the magnetic field $\mathsf{h}(\bm{r})$ to be statistically indefinite, rendering the correlation function Eq.~(\ref{Eq:3}) to be vanishing for $|\bm{r}^{\prime}-\bm{r}|\ne 0$.

The authors previously proposed that the 1D Berry phase term $\bm{\chi}$ universally induces an intermediate quasi-disordered phase between the ordered and disordered phases~\cite{zhaoTopologicalEffectAnderson2024c,shindou2025}. In the quasi-disordered phase, an exponential correlation length of the U(1)-phase correlation function is divergent along $\bm{\chi}$, while it is finite along the other directions.  According to the Coulomb loop gas model in Eq.~(\ref{Eq:4}), a vortex loop adds to the partition function a complex phase factor, and the phase is proportional to an area within the loop projected onto a plane perpendicular to $\bm{\chi}$. On the other hand, a polarized vortex loop---a vortex loop that is confined in a plane parallel to $\bm{\chi}$---does not induce any phase factor [Fig.~\ref{fig:loops}(b)]. Consequently, the complex phase factor selectively causes destructive interference among configurations of those vortex loops that have finite projections to the plane [Fig.~\ref{fig:loops}(a)], and the partition function in the ordered phase can be dominated by the polarized vortex loops, especially near a boundary of the ordered phase. In fact, the renormalization group analysis shows that the 1D Berry phase term suppresses the screening effect of unpolarized vortex loops, while the screening effect of the polarized vortex loops confines the larger vortex loops into planes parallel to $\bm{\chi}$, as well as stretches them along the $\bm{\chi}$ direction~\cite{shindou2025}. The proliferation of the polarized vortex loops makes the U(1)-phase correlation perpendicular to the $\bm{\chi}$ direction [Eq.~(\ref{Eq:3}) for $\bm{r}-\bm{r}^{\prime} \perp \bm{\chi}$] to be strongly disordered, while leaving intact the correlation function along the $\chi$ directions [Eq.~(\ref{Eq:3}) for $\bm{r}-\bm{r}^{\prime} \parallel \bm{\chi}$].  This may result in the emergence of the quasi-disorder phase next to the ordered phase, in which the exponential correlation length of the U(1)-phase correlation function is divergent only along $\bm{\chi}$. Such an intermediate quasi-disorder phase is consistent with the phenomenology of the quasi-localized phase in the chiral symmetric systems with the 1D weak band topology~\cite{xiaoAnisotropicTopologicalAnderson2023,zhaoTopologicalEffectAnderson2024c}. To promote this perspective, in the next section, we further elaborate on the analogy to magnetostatics and derive a dual representation of the 3D chiral NLSM. 

\section{Duality transformation}\label{sec:duality}
In the previous section, we showed that the saddle-point solution of the action is characterized by multiple vortex-loop configurations of an eigenvalue $e^{i\phi(\bm{r})}$ of the $Q$ field, and its associated eigenvector $\ket{p(\bm{r})}$ at vortex cores. In this section, we start from the partition function in the saddle-point manifold coordinated by these degrees of freedom, and derive its dual representation. Based on the dual model, we will argue in the next section that a mean field theory of the Anderson transition and the quasi-localization phenomena in the chiral symmetry class. For simplicity, we focus on the chiral unitary class (class AIII). Note that the duality transformation could be implemented in several different ways, depending on how to take a sum over vortex-loop configurations in the 3D space~\cite{savit1978,peskin1978,fradkin1978,herbutModernApproachCritical2007,kiometzisDualDescriptionSuperconducting1995}. In this paper, we place the continuous field theory into a 3D cubic lattice and derive a lattice version of the $\mathrm{U}(N)$ type-II superconductor model. 
%As in Ref.~\cite{kiometzisDualDescriptionSuperconducting1995}, one may also employ the particle-field duality and derive a $\mathrm{U}(N)$ generalization of the Anderson-Higgs model.

Key steps of the duality transformation can be summarized as follows. Following Ref.~\cite{konigMetalinsulatorTransitionTwodimensional2012,zhaoTopologicalEffectAnderson2024c}, we first rewrite the NLSM in terms of $\mathrm{u}(N)$ Lie algebra field $\bm{h}\equiv -iQ^{-1}\bm{\nabla} Q$, where a (generalized) rotation of ${\bm h}$ is locally constrained by vortex loop configurations. The constraint is then treated by a Lagrange multiplier $\bm{\Theta}$, which plays the role of dual vector potential field. The integration over $\bm{h}$ yields the dual theory, which still involves a sum over possible vortex loop configurations [Eq.~\eqref{eq:dual_continuous}]. To take such a sum specifically, we place the continuum theory onto a cubic lattice, and introduce a U(1) phase variable $\psi$ on each cubic lattice site. With a proper magnetic coupling between the U(1) variable and the dual vector potential, a partition function of the dual theory can be equated to a partition function of a $\mathrm{U}(N)$ generalization of the lattice superconductor [Eq.~\eqref{eq:dual_discretized}]. 

\subsection{Dual continuum theory}
The partition function in the saddle-point manifold involves the $Q(\bm{r})$-field integral, while the $Q(\bm{r})$ field is a $\mathrm{U}(N)$ matrix field with the nonlinear constraint $Q^{\dagger}Q=1$. To cope with the integral over such a curved manifold, we transform the integral variables into elements of the u($N$) Lie algebra via $\bm{h}\equiv -iQ^{-1}\bm{\nabla} Q$: the u($N$) field resides in a flat space~ \cite{konigMetalinsulatorTransitionTwodimensional2012, zhaoTopologicalEffectAnderson2024c}. In terms of the $h$ field, the action of the chiral NLSM reads
\begin{equation}\label{eq:AIII_NLSM_in_h}
\begin{aligned}
    S&=\int\frac{\mathrm{d}^3r}{8\pi}\sum_{\mu=x,y,z}\left(\sigma \Tr  h_\mu^2+c\Tr ^2h_\mu+i \chi_\mu\Tr h_\mu \right).
\end{aligned}
\end{equation} 
The saddle-point manifold is coordinated by configurations of vortex loops and $\ket{p(\bm{r})}$ at vortex cores. To relate the $\bm{h}$ field with these degrees of freedom, we introduce a field strength vector $F_\mu=\epsilon_{\mu\nu\rho}\nabla_\nu h_\rho + i\epsilon_{\mu\nu\rho}h_{\nu}h_{\rho}$~\cite{konigMetalinsulatorTransitionTwodimensional2012,zhaoTopologicalEffectAnderson2024c}. The field strength vector is directly given by the vortex loop configuration and $\ket{p(\bm{r})}$ at vortex cores (see Appendix \ref{app:saddle_point}), 
\begin{align}
    F_\rho(\bm{r})&=\epsilon_{\rho\mu\nu}\nabla_\mu \nabla_\nu \phi(\bm{r}) \!\ \ket{p(\bm{r})}\bra{p(\bm{r})} \nonumber \\
     &= 2\pi\sum_{j=1}^{n} \oint_{C_j} \mathrm{d}l_\rho\,\delta^3(\bm{r}-\bm{l})P(\bm{r}) \equiv J_{\rho}(\bm{r}), \label{eq:field-strength}
\end{align}
with $P(\bm{r}) \equiv \ket{p(\bm{r})}\bra{p(\bm{r})}$ and Eq.~(\ref{eq:winding}). Thus, by equating the right-hand side with $\mathrm{u}(N)$-valued vortex vector field $J_{\rho}(\bm{r})$ ($\rho=x,y,z$), we can write the partition function in the saddle-point manifold as follows, 
\begin{equation}
    Z=\int \mathrm{D}[\bm{h}] \int \mathrm{D}[\bm{J}]\,\delta\left[\bm{F}-\bm{J}\right]\,\exp\left(-S\right).
\end{equation}
Here, the integral of $\bm{h}$ is over the $\mathrm{u}(N)$ elements, while an integral of $\bm{J}$ comprises the sum over all possible multiple vortex-loop configurations, and the integral over $\ket{p(\bm{r})}$,
\begin{equation}
    \int \mathrm{D}[\bm{J}]\equiv\int\mathrm{D}[\phi_{\rm v}] \int \mathrm{D}[P]. 
\end{equation}
$\int \mathrm{D}[\phi_{\rm v}]$ is defined in Eq.~(\ref{Eq:5}), and $\int \mathrm{D}[P]$ stands for multiple integrals over $\ket{p(\bm{r})}$ at every $\bm{r}$ along vortex lines;
\begin{align}
\int \mathrm{D}[P] \equiv \prod_{\bm{r} \in \{C_j\}}\int \mathrm{d}\ket{p(\bm{r})}. 
\end{align}
The constraint $\bm{F}=\bm{J}$ can be implemented by an integral of a $\mathrm{u}(N)$-valued auxiliary vector field $\Theta_{\rho}(\bm{r})$ ($\rho=x,y,z$)
\begin{align}
 \delta[\bm{F}-\bm{J}]  =\int \mathrm{D}[\bm{\Theta}] \exp\left\{i\int \mathrm{d}^3r\,\Tr \left[\bm{\Theta}\cdot(\bm{F}-\bm{J})\right]\right\}. \label{eq:parameterization}
\end{align}
Thanks to the auxiliary field, the partition function becomes integrable over the $h$ field. The integration yields an action only of the auxiliary field $\bm{\Theta}(\bm{r})$ \cite{zhaoTopologicalEffectAnderson2024c},

\begin{align}
 & Z =  \int \mathrm{D}[\Theta] \!\ \!\ e^{-S_{\rm dual}}, \!\ \!\ \!\ S_{\text{dual}}=S_{\mathrm{u}(1)}+S_{\mathrm{su}(N)}+S_{t}, \label{eq:dual_continuous}\\
	& S_{\mathrm{u}(1)}=\frac{2\pi}{\sigma+N c} \int \mathrm{d}^3r\, \left(\bm{\nabla}\times\bm{\theta}^0-\frac{\sqrt{N}}{8\pi}\bm{\chi}\right)^2, \label{eq:dual_continuous_u1}\\
	& S_{\mathrm{su}(N)}=\frac{2\pi}{\sigma} \int \mathrm{d}^3r\,\sum_{a=1}^{N^2-1} \left(\bm{\nabla}\times \bm{\theta}^a\right)^2+\mathcal{O}(\sigma^{-2}), \label{eq:dual_continuous_suN}\\
	& S_t=-\int \mathrm{d}l\,t^{l} \int \mathrm{d}^3 R \int \mathrm{D}\Omega(\lambda) \int\mathrm{D}\ket{p(\bm{r}(\lambda))} \nonumber \\
    &\exp\left\{ 
    2\pi i \oint_C \mathrm{d}l_{\rho}  \left[\frac{\theta^0_{\rho}(\bm{l}) }{\sqrt{N}}
    + \sum^{N^2-1}_{a=1} \theta^a_{\rho}(\bm{l}) \braket{p(\bm{l}) |T^a|p(\bm{l})}\right]\right\} \label{eq:dual_continuous_Sy}.
\end{align}
Here $\bm{\theta}^0$ and $\bm{\theta}^a$ ($a=1,\cdots,N^2-1$) are real-valued expansion coefficients of the $\bm{\Theta}$ field in terms of the $\mathrm{u}(1)$ basis $T^0 \equiv 1/\sqrt{N}$ and $\mathrm{su}(N)$ bases $T^a$, respectively,
\begin{equation}
    \Theta_{\rho}(\bm{r})=\theta^0_{\rho}(\bm{r}) \!\ T^0+\sum_{a=1}^{N^2-1} \theta^a_{\rho}(\bm{r})\!\ T^a,
\end{equation}
with  $\Tr(T^0T^a)=0$, and $\Tr(T^a T^b)=\delta_{ab}$.

The dual theory \cref{eq:dual_continuous} respects a local $\mathrm{U}(1)$ gauge symmetry;
\begin{align}
\bm{\theta}^0(\bm{r}) \rightarrow \bm{\theta}^0(\bm{r}) + \bm{\nabla} \varphi(\bm{r}), 
\end{align}
with a rotation-free vector field $\bm{\nabla} \varphi(\bm{r})$. Thus, $\bm{\theta}^0(\bm{r})$, $S_{\mathrm{u}(1)}$, and $\sigma+N c$ can be interpreted as a magnetic gauge field, the magnetic part of the Maxwell action, and magnetic permeability, respectively. Similarly, $S_{\mathrm{su}(N)}$ can be also regarded as $\mathrm{SU}(N)$ generalization of the Maxwell action, while the dual theory does not have additional gauge symmetries associated with $\theta^a(\bm{r})$ ($a=1,\cdots,N$). In $S_{\mathrm{su}(N)}$, higher-order terms in $1/\sigma$, which encode interactions among S$\mathrm{U}(N)$ gauge fields, are neglected. The approximation may be justified near the metallic phase, where $\sigma$ is sufficiently large. $S_t$ comprises a sum over all possible configurations of {\it single} vortex loop, yielding highly nonlocal interactions among the gauge fields.   

The duality transformation also connects the correlation function, Eq.~(\ref{Eq:2}), to a ratio of the dual theory's partition functions with and without a pair of magnetic charges. As $\partial_{\mu}\phi = \Tr h_{\mu} \equiv h^0_{\mu} \sqrt{N}$, Eq.~(\ref{Eq:2}) can be expressed by a line integral of the u(1) component $h^0_{\mu}$,
\begin{align}
&\left\langle e^{i\phi(\bm{r}_1)-i\phi(\bm{r}_2)} \right\rangle = \frac{1}{Z} \int \mathrm{D}[\bm{h}]  \int \mathrm{D}[\bm{J}]\,\delta[\bm{F}-\bm{J}] \nonumber \\ 
&\times \exp\left[-S+i\sqrt{N} \int^{\bm{r}_1}_{\bm{r}_2}\mathrm{d}\bm{l}\cdot\bm{h}^0(\bm{l})\right] \nonumber \\
&= \frac{1}{Z} \int\mathrm{D}[\Theta] \,\exp\left[-S_{\mathrm{u}(1)}[v^{\rm D}(\bm{r})] - S_{{\rm su}(N)} - S_{t} \right], \label{eq:dual_ratio_continuum}
\end{align}
with
\begin{align}
&S_{\rm u(1)}[\bm{v}^{\rm D}(\bm{r})] = \nonumber \\
&\frac{2\pi}{\sigma+Nc} \int\mathrm{d}^3r\, 
\left[\bm{\nabla}\times\bm{\theta}^0-\frac{\sqrt{N}}{8\pi}\bm{\chi} + \sqrt{N} \bm{v}^{\rm D}(\bm{r})\right]^2. \label{eq:dual_continuous_u1_Dirac}
\end{align}
On the right-hand side of Eq.~(\ref{eq:dual_continuous_u1_Dirac}), $\bm{v}^{\rm D}(\bm{r})$ represents a quantized magnetic flux emanating from a magnetic monopole at $\bm{r}_1$ to a magnetic anti-monopole at $\bm{r}_2$, 
\begin{equation}
v^{\rm D}_{\mu} (\bm{r}) \equiv \int^{\bm{r}_1}_{\bm{r}_2} \mathrm{d}l_{\mu}\,\delta^3(\bm{r}-\bm{l}),\label{eq:dirac}
\end{equation}
namely, a Dirac string field~\cite{peskin1978,herbutModernApproachCritical2007}. Eq.~(\ref{eq:dual_ratio_continuum}) shows that the correlation function of the U(1) phase variables maps to the ratio between the dual theory's partition functions with and without the Dirac string field.
     
Eq.~(\ref{eq:dual_continuous}) suggests phase transitions induced by a change of the fugacity parameter, similar to those in the Coulomb loop gas models. Nonetheless, the non-local nature of the $S_t$ term hinders the systematic analysis of the dual continuum model. In the following subsection, we derive an alternative formulation of the dual action by discretizing the continuum theory on a cubic lattice. This approach introduces an emergent $\mathrm{U}(1)$ phase variable $\psi(\bm{r})$, yielding a fully local action in terms of $\bm{\Theta}(\bm{r})$ and $\psi(\bm{r})$.

\subsection{Dual lattice model}
\begin{figure}[htb]
    \centering
    \includegraphics[width=1.0\linewidth]{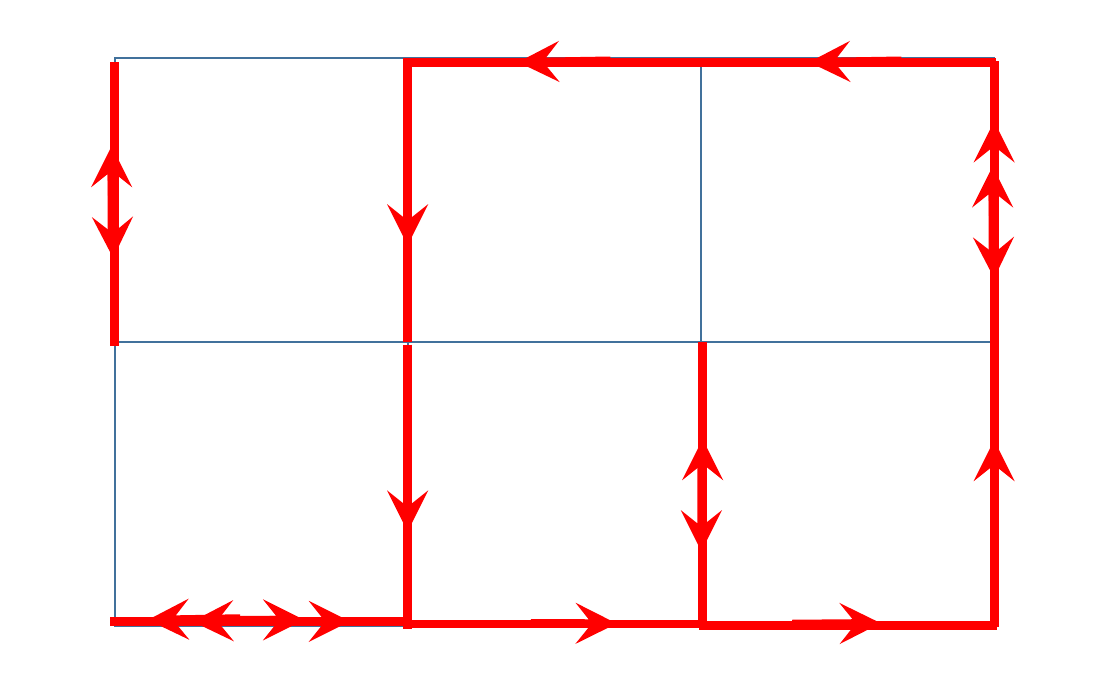}
    \caption{An example of closed-loop graphs with direction that have time-reversal pairs on the same links. For this closed-loop graph, the RHS of Eq.~(\ref{eq:10c}) gives a weight $t^{18}/(2! 2! 2!)$, while the LHS of Eq.~(\ref{eq:10c}) gives a weight $t^{18}$. The additional weight can be regarded as an additional fugacity for the higher vorticity.} 
    \label{fig:dual-lattice-model}
\end{figure}

To this end, let us begin with the partition function after the $h$-field integration,
\begin{align}
Z = &\int \mathrm{D}[\bm{\Theta}] e^{-S_{\rm u(1)} - S_{{\rm su}(N)}} \nonumber \\
   &\int \mathrm{D}[\phi_{\rm v}] \int \mathrm{D}[P]\,\ e^{- i \int \mathrm{d}r^3 \Tr [\bm{\Theta}(\bm{r})\cdot \bm{J}(\bm{r})]}. \label{eq:10a}
\end{align}
We first discretize the coupling term between $\bm{J}(\bm{r})$ and $\bm{\Theta}(\bm{r})$ on a cubic lattice, 
\begin{widetext}
\begin{align}
&\int \mathrm{D}[\phi_{\rm v}] \int \mathrm{D}[P]\, \ e^{- i \int \mathrm{d}^3r \Tr [\bm{\Theta}(\bm{r})\cdot\bm{J}(\bm{r})]} = 1 \!\ +  \nonumber \\
& \!\ \sum^{\infty}_{n=1} \frac{1}{n!} \prod^n_{j=1} \left(\int \mathrm{d}l_j \!\ t^{l_j} \int \mathrm{d}^3R_j \int_{\lambda \in [0,l_j]} \mathrm{D}\Omega_j(\lambda) \int_{r_j \in C_j} \mathrm{D}\ket{p(\bm{r}_j)}\right)  \exp\left[- 2\pi i \sum^n_{j=1}\oint_{C_j} \mathrm{d}l_{\mu}  \sum^{N^2-1}_{a=0} \theta^a_{\mu}(\bm{l}) \braket{p(\bm{l})|T^a|p(\bm{l})} \right]  \nonumber  \\ 
&\rightarrow 1 \!\  + \sum_{{\rm all} \ {\rm closed} \  {\rm loops}}  \,\ t^{L}   \prod_{m\in   {\rm all} \ {\rm the} \ {\rm loops} } \left(\int \mathrm{D}|p_m\rangle\right) \,\  \exp \left[-2\pi i  \sum_{n,\mu \in {\rm all} \ {\rm the} \ {\rm loops}}  \sum^{N^2-1}_{a=0} \theta^a_{n,\mu}\langle  p_n|T^a|p_n\rangle \right].
\label{eq:10b}
\end{align}
\end{widetext}
On the right-hand side (RHS), cubic lattice points are coordinated by an integer-valued 3D index $m \equiv (m_x,m_y,m_z)$. A closed vortex loop with direction is formed by connecting nearest-neighbor cubic lattice points. $\mathbb{CP}^{N-1}$ vector $|p_m\rangle$ is defined at cubic lattice point $m$. Magnetic gauge field $\theta^{a}_{m,\mu}$ $(a=0,1\cdots,N^2-1)$ resides on a nearest-neighbor link between $m$ and $m+\hat{\mu}$ ($\mu=x,y,z$), with a convention of an antisymmetric condition, $\theta^a_{m,-\mu}=-\theta^a_{m,\mu}$. Here $\hat{\mu}$ is the unit vector along the $+\mu$ axis. The first summation in the RHS (sum over all closed loops) is over all possible configurations of directed loops on the cubic lattice. In summation, two vortex lines can share links as well as lattice points. The second summation in the RHS (sum over $n,\mu \in$ all the loops) is over all links along directed loops. If loops have a link from $n$ to $n+\hat{x}$ and to $n-\hat{x}$, the summand includes $\theta^a_{n,x} \langle p_n|T^a |p_n\rangle$ and $-\theta^a_{n-\hat{x},x} \langle p_{n-\hat{x}} |T^a|p_{n-\hat{x}}\rangle$, respectively. The first product in the RHS (product over $m \in$ all the loops) is over all lattice points along directed loops. When a lattice point is traversed by the directed loops $q$ times, the product has multiple integrals over $q$ distinct $\mathbb{CP}^{N-1}$ vectors. Notably, the first summation in the RHS counts those directed loops with time-reversal pairs on the same links, where the path traverses the same links both in forward and backward directions (see Fig.~\ref{fig:dual-lattice-model}): the links with time-reversal pairs are regarded as having fine loop structures smaller than the lattice constant. $L$ in the RHS denotes the total number of links that constitute all the vortex loops. 

The sum over all directed-loop configurations on the cubic lattice can be taken by multiple integrals over U(1) phase variables $\psi_m$ defined on all cubic lattice sites, 
\begin{widetext}
\begin{align}
& 1 \!\  + \sum_{{\rm  all} \ {\rm closed} \  {\rm loops}}  \!\ t^{L} \,\ 
\prod_{m\in  {\rm all} \ {\rm the}  \ {\rm loops} }\left(\int \mathrm{D}|p_m\rangle \right)\, \
\exp \left[-2\pi i  \sum_{n,\mu \in {\rm all} \ {\rm the} \ {\rm loops}}  \sum^{N^2-1}_{a=0} \theta^a_{n,\mu}\langle  p_n|T^a|p_n\rangle \right] \nonumber \\
&\simeq  \prod_{m} \left(\int^{2\pi}_{0}\frac{\mathrm{d}\psi_m}{2\pi}\right) \sum^{\infty}_{L=0} \frac{t^L}{L!}
\left(\sum_{m} \sum_{\mu=x,y,z} \sum_{\sigma=\pm } \int \mathrm{d}|p_m\rangle \exp\left[i\sigma(\psi_{m+\hat{\mu}}-\psi_{m} + 2\pi \sum^{N^2-1}_{a=0} \theta^a_{m,\mu}\langle  p_m|T^a|p_m\rangle \right]\right)^L. \label{eq:10c}
\end{align}
\end{widetext}
On the RHS, the product and sum of $m$ are over all the cubic lattice points in the system. Suppose that the system has $M$ cubic lattice points. Then, $(\cdots)^L$ in the RHS can be expanded into $(M\times 3 \times 2)^L$ terms, each involving multiple integrals over $L$ distinct $\mathbb{CP}^{N-1}$ vectors (Here $M$, $3$ and $2$ come from the sums of $m$, $\mu$ and $\sigma$, respectively). Each term can be represented by a directed graph composed of nearest-neighbor links with arrows, where a link with an arrow from $m$ to $m+\hat{\mu}$ and to $m-\hat{\mu}$ represents a phase factor of  $\exp[i(\psi_{m+\hat{\mu}}-\psi_m+\cdots)]$ and $\exp[-i(\psi_{m}-\psi_{m-\hat{\mu}}+\cdots)]$, respectively. The multiple integrals over $\psi_m$ fields of all the lattice points eliminate most of the $(M\times 3 \times 2)^L$ terms, except for those terms represented by closed directed-loop graphs, e.g., Fig.~\ref{fig:dual-lattice-model}. Thereby, both the RHS and LHS of Eq.~(\ref{eq:10c}) count exactly the same closed-loop graphs with the same phase factors. The caveat is that while the LHS always gives $t^L$ for a closed directed-loop graph with $L$ links, the RHS gives an additional weight for some closed directed-loop graphs due to combinatorial factors. Namely, when a graph traverses the same link in the same direction $r$ times, the RHS gives the additional weight factor of $1/r!$ from the link. Since such an additional weight can be interpreted as additional core energies for vortex segments with {\it higher} vorticity, the RHS may as well be regarded as the partition function with a different lattice regularization. Given the RHS of Eq.~(\ref{eq:10c}) as the lattice model of $\int \mathrm{D}[\phi_{\rm v}]  \int \mathrm{D}[P] \!\ e^{- i \int \mathrm{d}^3r \Tr [\bm{\Theta}(\bm{r})\cdot \bm{J}(\bm{r})]}$, we finally obtain, 
\begin{align}
&\int \mathrm{D}[\phi_{\rm v}] \int \mathrm{D}[P] \!\ e^{- i \int \mathrm{d}^3r \Tr [\bm{\Theta}(\bm{r})\cdot \bm{J}(\bm{r})]} \nonumber \\
& \rightarrow \prod_{m} \left(\int^{2\pi}_{0}\frac{\mathrm{d}\psi_m}{2\pi}\right) \, \
\exp\Bigg[2t \sum_{m} \sum_{\mu=x,y,z}  \int \mathrm{d}|p_m\rangle \nonumber \\
& \hspace{0.1cm}\cos\bigg(\psi_{m+\hat{\mu}}-\psi_{m} + 2\pi  \sum^{N^2-1}_{a=0} \theta^a_{m,\mu}\langle  p_m|T^a|p_m\rangle \bigg)\Bigg]. 
\end{align}

Let us next discretize the Maxwell action on the cubic lattice~\cite{peskin1978,herbutModernApproachCritical2007}. On the lattice, the curl of the magnetic gauge field $(\bm{\nabla}\times \bm{\theta}^a)_{\overline{m},\mu}$ is defined on a dual lattice link connecting dual cubic lattice sites $\overline{m}$ and $\overline{m}+\hat{\mu}$. The curl represents an oriented sum of $\theta^a_{m,\nu}$ along edges of a plaquette intersected by the dual lattice link, e.g. 
\begin{align}
    (\bm{\nabla}\times\bm{\theta}^a)_{\overline{m},x}& \equiv -\theta^a_{\overline{m}+\hat{x}/2-\hat{y}/2+\hat{z}/2,y}+\theta^a_{\overline{m}+\hat{x}/2+\hat{y}/2-\hat{z}/2,z} 
    \nonumber \\
    &  \hspace{-0.5cm} +\theta^a_{\overline{m}+\hat{x}/2-\hat{y}/2-\hat{z}/2,y} - \theta^a_{\overline{m}+\hat{x}/2-\hat{y}/2-\hat{z}/2,z}.\label{eq:curl}
    \end{align}
To sum up, the partition function in the saddle-point manifold is mapped to the following dual lattice model,  
\begin{align}
&	Z \equiv \int \mathrm{D}[\bm{\Theta}] D[\psi]  \!\ \exp [-S_{\mathrm{u}(1)}-S_{\mathrm{su}(N)}-S_t], \label{eq:dual_discretized}\\
	& S_{\mathrm{u}(1)}=\frac{2\pi\Lambda^{-3}}{\sigma+Nc} \sum_{\overline{m},\mu}\left[\Lambda \left(\bm{\nabla}\times\bm{\theta}^0\right)_{\overline{m},\mu}-\frac{\sqrt{N}}{8\pi}\chi_\mu\right]^2, \label{eq:dual_discretized_u1}\\
	& S_{\mathrm{su}(N)}=\frac{2\pi\Lambda^{-1}}{\sigma} \sum_{\overline{m},\mu}\sum_{a=1}^{N^2-1}\left(\bm{\nabla}\times\bm{\theta}^a\right)^2_{\overline{m},\hat{\mu}}, \label{eq:dual_discretized_suN}\\
	& S_t=-2 t\sum_{m,\mu}\int \mathrm{d} \!\ |p_{m}\rangle  \!\ \cos\Bigg[\psi_{m+\hat{\mu}}-\psi_m \nonumber \\
  &\hspace{3cm}  +2\pi \Lambda^{-1} \sum^{N^2-1}_{a=0} \theta^a_{m,\mu} \langle p_m | T^a |p_m\rangle \Bigg] \label{eq:dual_discretized_Sy}.
\end{align}
Here, we recover a lattice constant $\Lambda^{-1}$ of the cubic lattice for booking purposes. As in the dual continuum theory, the correlation function of the U(1) phase variable, Eq.~(\ref{Eq:2}), also maps to the ratio between the dual theory's partition functions with and without the Dirac string field $\bm{v}^{\rm D}$~\cite{peskin1978,herbutModernApproachCritical2007},
\begin{align}
\left\langle e^{i[\phi(\bm{r}_1)-\phi(\bm{r}_2)]}\right\rangle = \frac{Z[\bm{v}^{\rm D}]}{Z[0]}, \label{eq:dual_ratio}
\end{align}
and
\begin{align}
&Z[\bm{v}^{\rm D}] = \nonumber \\
& \ \int \mathrm{D}[\bm{\Theta}] \mathrm{D}[\psi]  \!\ \exp\left\{-S_{\mathrm{u}(1)}[\bm{v}^{\rm D}]-S_{\mathrm{su}(N)}-S_t\right\}, \label{eq:dual_discretized_vD}\\
	 & S_{\mathrm{u}(1)}[\bm{v}^{\rm D}] = \nonumber \\ 
     &  \frac{2\pi\Lambda^{-3}}{\sigma+Nc} \sum_{\overline{m},\mu}\left[\Lambda\left(\bm{\nabla}\times\bm{\theta}^0\right)_{\overline{m},\mu}-\frac{\sqrt{N}}{8\pi}\chi_\mu + \sqrt{N} v^{\rm D}_{\overline{m},\mu}\right]^2. \label{eq:dual_discretized_u1_vD} 
\end{align}
On the lattice, the Dirac string field $v^{\rm D}$ is depicted by a directed line from dual cubic lattice site $\overline{m}_1$ (that corresponds to $\bm{r}_1$) to $\overline{m}_2$ (that corresponds to $\bm{r}_2$). The open line is formed by connecting nearest-neighbor dual-cubic-lattice links. $v^{\rm D}_{\overline{m},\mu}$ in Eq.~(\ref{eq:dual_discretized_u1_vD}) takes $+1$ and $-1$, when a link from $\overline{m}$ to $\overline{m}+\hat{\mu}$ is traversed by the directed line in the same and opposite direction, respectively. Otherwise, $v^{\rm D}_{\overline{m},\mu}=0$.

The dual lattice model for $N=1$ portrays the magnetostatics of a U(1) type-II superconductor subjected to an external magnetic field along $\bm{\chi}$~\cite{peskin1978,dasgupta1981,herbutModernApproachCritical2007}. In the model, $\psi_m$ represents the U(1) phase of the superconducting order parameter on the cubic lattice site $m$. The U(1) phase $\psi_m$ couples with the U(1) gauge field $\bm{\theta}^0_m$ through the magnetic coupling. In the dual model, the fugacity parameter $t$ becomes the Josephson coupling constant, and the 1D weak topological term $\bm{\chi}$ becomes the external magnetic field. When $\chi=0$, the disordered phase, characterized by a divergent fugacity parameter $t$, maps to the superconducting phase where the large Josephson coupling $t$ breaks the global U(1) gauge symmetry spontaneously, and the U(1) gauge field acquires a finite mass via the Anderson-Higgs mechanism~\cite{anderson1963,higgs1964}. Conversely, the ordered phase, with vanishing fugacity parameter $t$,  maps to the normal phase, where the fluctuation of the U(1) gauge field, favored by the entropic effect but constrained by the Maxwell action, completely suppresses the superconducting order.  

A similar mapping of the phase diagram is also expected for general $N$, as the models share the same local U(1) gauge symmetry,
\begin{align}
& \psi_m \rightarrow \psi_m - \varphi_m,  \\
& \theta^0_{m,\mu} \rightarrow \theta^0_{m,\mu} 
+ \frac{\sqrt{N}}{2\pi\Lambda^{-1}}(\varphi_{m+\hat{\mu}}-\varphi_m), 
\end{align}
In the next section, we first discuss a mean-field picture of the large-$t$ (superconducting, Meissner) phase for general $N$. 

\section{Anderson localization in chiral unitary class}\label{sec:anderson}
When the large $t$ induces the spontaneous symmetry breaking of the global U(1) gauge symmetry, an associated gapless Goldstone mode is absorbed by the U(1) gauge field, and the gauge field acquires a finite mass. To see this mass acquisition, let us expand the Josephson coupling term up to the second order in the gauge field fluctuation $\delta \bm{\Theta}(\bm{r}) = \sum^{N^2-1}_{a=0} \delta \bm{\theta}^a(\bm{r}) T^a$,
\begin{align}
        \delta S&=2\pi\int \mathrm{d}^3r\,\left[\frac{\left(\bm{\nabla}\times\delta\bm{\theta^0}\right)^2}{\sigma+Nc}+\sum_{a=1}^{N^2-1}\frac{\left(\bm{\nabla}\times\delta\bm{\theta}^a\right)^2}{\sigma}\right]\\
        &+4\pi^2 t \Lambda^{3} \int \mathrm{d}^3r\int |p(\bm{r})\rangle \,\Tr ^2\left[\Lambda^{-1}\delta\bm{\Theta}(\bm{r}) P(\bm{r})\right],
\end{align}
with $\Lambda^{-3}\sum_{m} \rightarrow \int\mathrm{d}^3 r$. In terms of a formula~\cite{konigMetalinsulatorTransitionTwodimensional2012,zhaoTopologicalEffectAnderson2024c}
\begin{equation}
    \begin{aligned}
        &\int \mathrm{d}|p\rangle \,\Tr (A P)\Tr (B P)\\
        &=\frac{\pi^{N-1}}{N(N+1)\Gamma(N)}\left[\Tr (AB)+\Tr (A)\Tr (B)\right],
    \end{aligned}
\end{equation}
the integral over $p(\bm{r})$ yields a decoupling among the $N^2$ gauge fields, 
\begin{align}
&\delta S=2\pi\int \mathrm{d}^3r\left[\frac{(\bm{\nabla}\times\delta \bm{\theta}^0)^2}{\sigma+Nc}+\frac{2\pi^{N}t\Lambda}{N\Gamma(N)} (\delta \bm{\theta}^0)^2 \right] \nonumber \\
&+2\pi\int \mathrm{d}^3r\sum_{a}\left[\frac{(\bm{\nabla}\times\delta \bm{\theta}^0)^2}{\sigma}+\frac{2\pi^{N}t\Lambda}{N(N+1)\Gamma(N)}(\delta \bm{\theta}^a)^2\right]. \label{eq:action_for_SC}
\end{align}
Notably, both U(1) and S$\mathrm{U}(N)$ gauge fields acquire finite mass, while the two masses converge to the same finite constant in the replica limit $(N\rightarrow 0)$. The mass of the U(1) gauge field gives an energy penalty to the partition function with the Dirac string field, $Z[v^{\rm D}]$,  in comparison to the partition function without the Dirac string field, and the energy penalty is proportional to the distance between the magnetic monopole and antimonopole. Thus, the U(1) phase correlation function given by Eq.~(\ref{eq:dual_ratio}) decays exponentially in the distance $|\bm{r}_1-\bm{r}_2|$ for general integer $N$. In the limit of $N\rightarrow 0$, such a short-ranged correlation of the U(1) phase variables is consistent with the Anderson insulator in the chiral NLSM. When the mass is absent ($t=0$), the action is given only by the Maxwell term, where an energy penalty that $Z[v^{\rm D}]$ pays relative to $Z[0]$ in Eq.~(\ref{eq:dual_ratio}) is given by the $1/r$ Coulomb interaction between the magnetic charges. Thereby, the correlation function remains finite even for larger $|{\bm r}_1-{\bm r}_2|$, which is consistent with the diffusive metal phase in the chiral NLSM. 

In the next section, we will develop a mean-field theory of the superconducting phase of the $\mathrm{U}(N)$ lattice superconductor model for general $N$. We will show that the theory in the replica limit $(N\rightarrow 0)$ has a phase transition from a large-$t$ phase (superconducting, Meissner / Anderson localized) phase to a small-$t$ ordered phase (normal, Maxwell /  diffusive metal) phase. 

\subsection{Variational analysis}
To gain a mean-field gap equation for the Higgs mass for the gauge field, we use a variational method and study the continuum limit of the action $S_{\rm u(1)}+S_{{\rm su}(N)}+S_t$:
\begin{align}
S =&  \int \mathrm{d}^3r\left\{2\pi \left[\frac{\left(\bm{\nabla}\times\bm{\theta}^0
    \right)^2}{\sigma+Nc}+\sum_{a=1}^{N^2-1} 
    \frac{\left(\bm{\nabla}\times\bm{\theta}^a\right)^2}{\sigma}\right]\right. \nonumber \\
   &\left. - 2t \Lambda^{3}\sum_\mu \cos\left[2\pi\Lambda^{-1}\sum^{N^2-1}_{a=0}\theta^{a}_{\mu}(\bm{r}) \langle p(\bm{r})|T^a|p(\bm{r})\rangle\right]\right\}.
\end{align}
Here the spatial gradient $\nabla_{\mu}\psi(\bm{r})$ of the superconducting phase is already absorbed into the U(1) gauge field, $\theta^{0}_{\mu}(\bm{r}) \rightarrow \theta^{0}_{\mu}(\bm{r}) + \nabla_{\mu}\psi(\bm{r})$. Based on a variational principle, we search for a trial action $\overline{S}$ that minimizes a variational free energy $F \equiv \braket{S-\overline{S}}_{\overline{S}}-\ln\overline{Z}$, where the average $\braket{\cdots}_{\overline{S}}$ is taken with a weight of $\exp(-\overline{S})$, and $\overline{Z}$ is the partition function of the trial action. From Eq.~(\ref{eq:action_for_SC}) in the replica limit, we assume the following form of the trial action,  
\begin{align}
    \overline{S}=2\pi\int \mathrm{d}^3r\Bigg\{&\left[\frac{\left(\bm{\nabla}\times\bm{\theta}^0
    \right)^2}{\sigma+Nc}+m^2\left(\bm{\theta}^0\right)^2\right]\\
    +\sum_{a=1}^{N^2-1}&\left[\frac{\left(\bm{\nabla}\times\bm{\theta}^a\right)^2}{\sigma}+m^2\left(\bm{\theta}^a\right)^2\right]\Bigg\}, \label{Eq:trial}
\end{align}
where the Higgs mass $m$ plays the role of a variational parameter. $F$ as a function of $m$ can be calculated for general $N$. Notably, $F$ thus calculated goes to zero in the replica limit, so that $f \equiv \lim_{N\rightarrow 0}F/(NV)$ corresponds exactly to $-\langle \ln \mathcal{Z}\rangle_{\rm dis}/V$ in Eq.~(\ref{eq:replica}) (see \cref{app:variational_method}).  $f$ is calculated as a function of $m$,
\begin{widetext}
    \begin{align}
        f(m)=-\frac{6t \Lambda^{3}}{\pi}\exp\left\{- \frac{\Lambda}{48}\frac{\Lambda^3+m\left[3\sqrt{\sigma}\Lambda^2+m\left(2c\Lambda+5\sigma \Lambda +3m\sigma^{\frac{3}{2}}\right)\right]}{m^2\left(\Lambda+\sqrt{\sigma}m\right)^3}\right\} +\frac{1}{8\pi}\frac{c\sqrt{\sigma}m^3 \Lambda^3}{\left(\Lambda+m\sqrt{\sigma}\right)^3}. \label{mf-energy}
    \end{align}
    \end{widetext}
where $\Lambda$  is the UV cutoff for 3D momentum integrals in the calculation. We use a soft momentum cutoff for the 3D integrals; $\Lambda^{-1}$ can be regarded as a lattice constant.

\begin{figure*}[htb]
    \centering
    \includegraphics[width=1.0\linewidth]{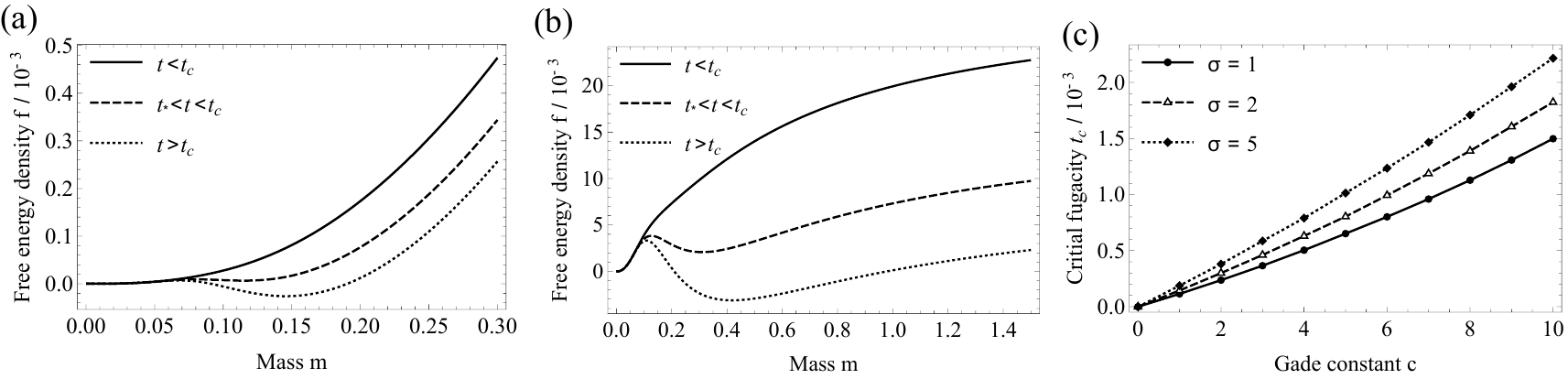}
    \caption{Variational free energy density $f/\Lambda^3$ as a function of the Higgs mass $m/\sqrt{\Lambda}$ at (a) $\sigma/\Lambda=c/\Lambda=1$, and (b) $\sigma/\Lambda=c/\Lambda=80$ (for the $\chi=0$ case). 
    $f$ has a single minimum at $m=0$ for $t<t_*$, two local minima at $m=0$ and at $m\ne 0$ for $t_*<t$. The two minima of $f$ are degenerate at $t=t_c$, while for $t_c<t$, the global minimum of $f$ is at $m\ne 0$. The critical fugacities for $\sigma/\Lambda=c/\Lambda=1$, and $\sigma/\Lambda=c/\Lambda=80$ are $t_c=1.15 \times 10^{-4}$, and $1.37\times 10^{-2}$, respectively. (c) Critical fugacity $t_{\text{c}}$ for different conductivity $\sigma/\Lambda$ and Gade constant $c/\Lambda$. In these figures, we choose $\Lambda^{-1}=1$ for simplicity.}
    \label{fig:mf_nt}
\end{figure*}

The first and second terms of $f(m)$ decrease and increase monotonically in increasing $m$, respectively, so that a competition between the two terms leads to a phase transition between the $m\ne 0$ phase [Anderson localized phase] and  $m=0$ phase [diffusive metal phase]. In fact, the entropic effect associated with the gauge field fluctuations, that comes from $-\ln \overline{Z}$, is encoded in an increasing function of $m$ in the second term, while internal energy that comes from $\langle S-S_{0}\rangle_{\overline{S}}$ decreases in $m$, e.g. the first term. The critical value $t_c$ of $t$ is calculated for different values of $\sigma$ and $c$ [Fig.~\ref{fig:mf_nt}(c)], and the free energy density $f$ is plotted as functions of $m$  for $t>t_c$ and  $t<t_c$. When the fugacity parameter $t$ is below the critical value $t_{\text{c}}$, the global minimum of $f$ is at $m=0$; when $t$ is above $t_{\text{c}}$,  the global minimum of $f$ appears at $m\ne 0$. Notably, the function $f$ at $t=t_{\text{c}}$ has two minima at $m=0$ and at $m\ne 0$ [Fig.~\ref{fig:mf_nt}(a,b)].

The Anderson transition in the 3D chiral unitary class is a second-order phase transition~\cite{wang2021,xiaoAnisotropicTopologicalAnderson2023}, and the first-order transition suggested by the two minima of the trial function is an artifact of the variational approximation. The approximation may underestimate the fluctuations of the superconducting phase and magnetic gauge fields. A similar first-order phase transition was also predicted by a mean-field analysis of the Anderson-Higgs model~\cite{herbutModernApproachCritical2007}. Note also that the critical fugacity $t_c$ at $\sigma\ne 0$ vanishes in the limit of $c=0$ [Fig.~\ref{fig:mf_nt}(c)]. This stems from Eq.~(\ref{mf-energy}) whose second term vanishes completely at $c=0$. On the other hand, it is physically likely that $t_c \ne 0$ at $\sigma\ne 0$ and $c=0$, because the phase stiffness of the U(1) phase variable $\phi$ is given by $\sigma+c$ [see Eq.~(\ref{Eq:1})]. The discrepancy might be resolved by including the interaction term among SU$(N)$ gauge fields, which may add other polynomials of $m$ with different coefficients in Eq.~(\ref{mf-energy}). 

\section{Quasilocalization in chiral unitary class}\label{sec:quasilocalization}
In the presence of $\chi$, the lattice model of U(1) type-II superconductors exhibits a two-step phase transition, where an intermediate mixed phase emerges between the normal (Maxwell) and superconducting (Meissner) phases. In this mixed phase, the system is permeated by a dense array of magnetic flux lines aligned along the $\bm{\chi}\parallel z$ direction, spaced at finite intervals in the $x$-$y$ plane~\cite{blatter1994,brandt1995,zeldov1995}. 
  
In the mixed phase, the monopole-field correlation function, Eq. (\ref{eq:dual_ratio}), remains finite for large $|\bm{r}_1-\bm{r}_2|$ in the parallel geometry ($\bm{r}_1-\bm{r}_2 \parallel z$ axis), while it decays exponentially with the distance in the perpendicular geometry ($\bm{r}_1-\bm{r}_2 \perp z$ axis). This behavior arises because the Dirac string field $\bm{v}^{\rm D}$ emanating from the magnetic monopole to antimonopole in the parallel geometry is trapped inside one of the magnetic flux lines near $\bm{r}_1$ and $\bm{r}_2$. As a result, the additional energy penalty that $Z[\bm{v}^{\rm D}]$ incurs relative to $Z[0]$ is given only by the Coulomb interaction between the two magnetic charges, which vanishes as $|\bm{r}_1-\bm{r}_2|\rightarrow \infty$. In contrast, the Dirac string field in the perpendicular geometry is inevitably exposed to some superconducting region, and the region is proportional to $|\bm{r}_1-\bm{r}_2|$. Consequently, the additional energy penalty scales linearly with $|\bm{r}_1-\bm{r}_2|$, leading to an exponential decay of the U(1) phase correlation function Eq.~(\ref{eq:dual_ratio}) along the perpendicular direction.

A similar phase transition from the normal phase to the intermediate mixed phase is also expected in the $\mathrm{U}(N)$ type-II superconductors for general $N$. When $t$ is small or both $\sigma$ and $c$ are large, the superconducting phase is disordered, and the Josephson couplings vanish completely due to the fluctuations of the $\psi$ field and gauge fields. When $t$ is large or both $\sigma$ and $c$ are small, the internal energy dominates the entropic effect, and the system chooses to take the saddle-point solution(s) of Eq.~(\ref{eq:dual_discretized}). The global energy minimum of $S_{\rm u(1)}$ is given by  
\begin{equation}
    \bar{\theta}^0_{m,\mu}=\Delta_{\mu}\varphi_{m}+\frac{\sqrt{N}}{8\pi}A_{m,\mu}. \label{eq:saddle_1}
\end{equation}
Here $\Delta_{\mu} \varphi_m \equiv \varphi_{m+\hat{\mu}}-\varphi_m$ is a longitudinal (rotation-free) component of the U(1) gauge field $\bar{\theta}^0_{m,\mu}$, and an external magnetic gauge potential $\bm{A}$ satisfies $(\bm{\nabla}\times\bm{A})_{\overline{m},\hat{\mu}}=\chi/\Lambda \, \delta_{\mu,z}$. The energy minimum of $S_{{\rm su}(N)} $ can be achieved by the uniform $\mathrm{su}(N)$ gauge field,  
\begin{equation}
    \bar{\theta}^{a\ne 0}_{m,\hat{\mu}}=0. \label{eq:saddle_2}
\end{equation}
With Eqs.~(\ref{eq:saddle_1},\ref{eq:saddle_2}), $S_t$ is minimized by   
\begin{align}
    \bar{\psi}_{m+\hat{\mu}}-\bar{\psi}_m+\frac{2\pi \Delta_{\mu}\varphi_{m}}{\sqrt{N}\Lambda}+\frac{A_{m,\hat{\mu}}}{4\Lambda}=2\pi \mathbb{Z}. \label{eq:saddle_3}
\end{align}
For $\chi=0$, $\overline{\psi}_m=-2\pi \varphi_m/(\sqrt{N}\Lambda)$ realizes Eq.~(\ref{eq:saddle_3}) for all $\mu=x,y,z$ and $m$. For general values of $\chi$, however, Eq.~(\ref{eq:saddle_3}) cannot be simultaneously satisfied for all links. Namely, the oriented sum of Eq.~(\ref{eq:saddle_3}) around a cubic-lattice plaquette in the $x$-$y$ plane yields $\chi/(4\Lambda^2)$ from the right-hand side, while it gives $2\pi$ times integers from the left-hand side. On the other hand, Eq.~(\ref{eq:saddle_3}) for $\mu=z$ can be simultaneously satisfied for all $m$, as $A_{m,z}=0$. Accordingly, the system may choose a phase where the U(1) superconducting phase variable preserves its coherence along $\bm{\chi} \parallel z$, while the Josephson couplings within the $x$-$y$ plane are destroyed completely by the fluctuation of the $\psi$ field. In such a phase, the cos terms along $z$ links can be replaced by a quadratic fluctuation around its saddle point, yielding the following effective action, 
{\small
\begin{align}
    &S_{\rm quasi}=2\pi\int \mathrm{d}^3r\left[\frac{(\bm{\nabla}\times\delta\bm{\theta}^0)^2}{\sigma+Nc}+\frac{2t\pi^{N}\Lambda}{N\Gamma(N)}\left(\delta\theta^0_z\right)^2\right]\\
    &+2\pi\int \mathrm{d}^3r\sum^{N^2-1}_{a=1}\left[\frac{(\bm{\nabla}\times\delta\bm{\theta}^a)^2}{\sigma}+\frac{2t\pi^{N}\Lambda}{N(N+1)\Gamma(N)}\left(\delta\theta^a_z\right)^2\right], \label{eq:action_fluc_topological}
\end{align}
}
with $\bm{\theta}^a = \overline{\bm{\theta}}^a+ \delta \bm{\theta}^a$. Notably, only the longitudinal component $\delta\theta^a_z$ of the gauge fields acquires a finite mass, while the transverse components $\delta\theta^a_{\mu}$ ($\mu=x,y$) do not. In such a phase, the introduction of the Dirac string field along $x$ or $y$ directions incurs an energy penalty that is proportional to the string length, resulting in an exponential decay of the U(1) phase correlation along the $x$-$y$ direction. Conversely, the Dirac string field along $\bm{\chi} \parallel z$ is free from the mass so that the U(1) phase variable retains its coherence along $\bm{\chi}$. This extremely anisotropic U(1) phase coherence is consistent with the phenomenology of the quasi-localized phase in the chiral symmetric systems with the 1D weak band topology~\cite{xiaoAnisotropicTopologicalAnderson2023, zhaoTopologicalEffectAnderson2024c}.  

The above mean-field description also suggests that the intermediate quasi-disordered phase in the Coulomb loop gas model with finite $\chi$ must be characterized by divergent fugacity parameter $t_z$ for vortex loop segment polarized along $\bm{\chi} \parallel z$, and vanishing fugacity parameter $t_{\perp}$ for vortex loop segment polarized along the other directions. In fact, the mean-field action is consistent with a superconductor model with spatially anisotropic Josephson couplings, $t_{z} \ne t_{\perp} =t_x=t_y$, 
\begin{widetext}
   \begin{align}
   &Z = \int \mathrm{D}[\Theta] \int \mathrm{D}[\psi]  \, \ \exp \left[-S_{\rm u(1)} - S_{{\rm su}(N)}\right]
   \nonumber \\ 
   &\sum^{\infty}_{L=0} \frac{1}{L!} 
\left(\sum_{m} \sum_{\mu=x,y,z} \sum_{\sigma=\pm } \int \mathrm{d}|p_m\rangle\,\ t_{\mu} \exp\left[i\sigma(\psi_{m+\hat{\mu}}-\psi_{m} + 2\pi \Lambda^{-1}\sum^{N^2-1}_{a=0} \theta^a_{m,\mu}\langle  p_m|T^a|p_m\rangle \right]\right)^L. 
   \end{align}
\end{widetext}
Thereby, Eq.~(\ref{eq:action_fluc_topological}) can be regarded as a mean-field action for a phase with divergent $t_{z}$, and vanishing $t_{\perp}$. Notably, in the Coulomb loop gas model, $\ln t_{z}$ and $\ln t_{\perp}$ are nothing but chemical potentials of the vortex loop segment polarized along $\bm{\chi}\parallel z$ and the loop segment polarized along the other two directions, respectively. Thus, a transition from ordered (diffusive metal) to quasidisordered (quasilocalized) phase in the Coulomb loop gas model can be regarded as a topological transition driven by the spatial proliferation of {\it polarized} vortex loops.   

\section{Summary}\label{sec:discussion}
In this paper, we argue that vortex loop excitations drive the 3D Anderson transition in chiral symmetry classes, where an eigenvalue of the NLSM matrix-formed field variable $Q$ permits the vortex-loop solutions as saddle-point configurations of the NLSM. Using an analogy to 3D magnetostatics in the presence of quantized electric current loops, we argue that the Anderson transition in chiral symmetry classes is driven by a condensation of the vortex loop excitations (quantized electric current loops). To elaborate on this picture, we further introduce a cubic-lattice model of $\mathrm{U}(N)$ type-II superconductors as a dual representation of the NLSM in the chiral unitary class. By analyzing this lattice model by a variational method, we construct a mean-field theory of the 3D Anderson transition in the chiral unitary class. In the mean-field description, the metal phase with gapless diffusion mode becomes a normal (Maxwell) phase, where the Maxwell action controls the fluctuation of the gapless magnetic gauge field. Meanwhile, the localized phase is described as a superconducting (Meissner) phase with broken global U(1) gauge symmetry, where the gauge field acquires a finite mass via the Anderson-Higgs mechanism. Our dual representation also demonstrates that the 1D weak topology term in the 3D chiral NSLM becomes an external magnetic field in the 3D type-II superconductor model, and that the quasi-localized phase in chiral symmetry classes can be a similar spatially inhomogeneous phase as the mixed phase in the 3D superconductor. Thereby, the 3D localized bulk system is permeated by an array of 1D quasi-conducting regions aligned along the ${\bm \chi}\parallel z$ direction, while the 1D conducting regions are spaced by a finite distance within the $x$-$y$ plane.    

\begin{acknowledgements} 
We thank Lingxian Kong, Yeyang Zhang, and Zhenyu Xiao for the discussions. The work was supported by the National Basic Research Programs of China (No. 2024YFA1409000) and the National Natural Science Foundation of China (No. 12074008 and No. 12474150).
\end{acknowledgements}

\onecolumngrid
\appendix
\section{Derivation of the nonlinear sigma model in chiral unitary class}\label{app:derive_NLSM}
%In this section, we outline the derivation of NLSM for the chiral symmetry class. %For example, we consider class AIII.}
A Hamiltonian $H$ with chiral symmetry, $\sigma_z H \sigma_z = - H$, is given by a two-by-two block off-diagonal form, $H=(h_0+V)\sigma_{+} + (h^\dagger_0+V^\dagger)\sigma_{-}$, where $h_0$ and $V$ are $M\times M$ non-Hermitian matrices. $h_0$ is a disorder-free tight-binding matrix. $V$ is a diagonal matrix whose diagonal elements $V_r$ are random complex numbers obeying the Gaussian distribution function with $\langle V_{r}\rangle_{\rm dis}=0$ and $\langle V_{r}^*V_{r^\prime}\rangle_{\rm dis}=\gamma^2\delta_{rr^\prime}$. A Green's function $G(\omega;r,r^{\prime})=\langle r|(\omega-H)^{-1}|r^{\prime}\rangle$ and its products are generated from a partition function,
\begin{equation}
    {\cal Z}=\int \mathrm{D}[\psi]\,\exp\left[-\psi^\dagger(\omega-H)\psi\right]
    =\int \mathrm{D}[\psi]\,\exp\left[-\psi^\dagger(\omega-H)\sigma_x\psi\right],
\end{equation}
with $\psi \equiv (\psi_{+},\psi_{-})^T$ and $\psi^{\dagger}\equiv (\psi^{\dagger}_{+},\psi^{\dagger}_{-})$. Here $\pm$ is the chiral index associated with the two-by-two block-off diagonal structure~\cite{altlandTopologyAndersonLocalization2015,konigMetalinsulatorTransitionTwodimensional2012,zhaoTopologicalEffectAnderson2024c}. The right-hand side is due to $|\det\sigma_x|=1$. The Green's function and its product are obtained from derivatives of the partition function $\cal Z$ with respect to a source field $J$, e.g. 
\begin{equation}
    G(\omega;r,r^{\prime})=\frac{\partial\ln {\cal Z}}{\partial J_{r,r^{\prime}}}.
\end{equation}
As the derivatives and the disorder average commute with each other, we have only to calculate $\langle \ln  {\cal Z}\rangle_{\rm dis}$. To calculate $\langle \ln{\cal Z}\rangle_{\rm dis}$, we use the replica trick,  
\begin{equation}
    \ln {\cal Z}=\lim_{N\to 0}\frac{{\cal Z}^N-1}{N},
\end{equation}
where ${\cal Z}^N$ is a partition function of $N$ replicated fermion field $\psi_m\equiv (\psi_{m,+},\psi_{m,-})^T$ ($m=1,\cdots,N$). The Gaussian average of ${\cal Z}^N$ induces an attractive interaction among the replicated $\psi$-fields, 
\begin{equation}
    \braket{{\cal Z}^N}_{\text{dis}}=\int \mathrm{D}[\psi]\,\exp\left[-\sum^N_{n=1}\psi_n^\dagger(\omega-H_0)\sigma_x\psi_n+\frac{\gamma^2}{2}\sum^N_{m,n=1} \sum^M_{r=1} \psi_{m,+,r}^*\psi_{m,+,r}\psi_{n,-,r}^*\psi_{n,-,r}\right],
\end{equation}
with $H_0  =  h_0\sigma_{+} + h^\dagger_0\sigma_{-}$. $H_0$ is diagonal in the replica index ($n$) and independent of the replica index $n$.  

\subsection{Self-consistent Born approximation}
The attractive interaction is further decomposed by the use of the Hubbard-Stratonovich (HS) transformation,
\begin{align}\label{eq:6}
&\exp\left[\frac{\gamma^2}{2}\sum_{m,n,r}\psi_{m,+,r}^*\psi_{m,+,r}\psi_{n,-,r}^*\psi_{n,-,r}\right] \nonumber \\
  & = \int \mathrm{D}[Q]\,\exp\left[-\frac{1}{2\gamma^2}\Tr (Q^\dagger Q)+i\left(\psi^\dagger_{+}Q \psi_{-}+\psi^\dagger_{-}Q^\dagger \psi_{+}\right)\right].
\end{align}
Here, the HS field ($Q$ field) takes the form of an $N$ by $N$ matrix. The $Q$ field is diagonal in the lattice-site index $r$, while it is dependent on $r$. The partition function of the $N$ replicated fields is given by 
\begin{equation}\label{eq:A7}
    \braket{{\cal Z}^N}_{\text{dis}}=\int \mathrm{D}[Q,\psi]\,\exp\left[-\frac{1}{2\gamma^2}\Tr \left(Q^\dagger Q\right)-\psi^\dagger
    \begin{pmatrix}
        \omega-iQ & -h_0 \\
        -h^\dagger_0 & \omega-iQ^\dagger
    \end{pmatrix}
    \sigma_x\psi
    \right].
\end{equation}
An integration of the fermion $\psi$-field yields the partition function for the auxiliary $Q$ field,
\begin{align}
    \braket{{\cal Z}^N}_{\text{dis}} &=\int \mathrm{D}[Q]\,\exp\left[-\frac{1}{2\gamma^2}
    \Tr \left(Q^\dagger Q\right)+\Tr \ln
    \begin{pmatrix}
        \omega-iQ & -h_0 \\
        -h^\dagger_0 & \omega-iQ^\dagger
    \end{pmatrix}
    \right], \nonumber \\
    & \equiv\int \mathrm{D}[Q] \exp \left[-S[Q,\omega]\right]. 
\end{align}
Here, the trace on the right-hand side is taken over the chiral index ($\pm$), replica index ($m,n$), and the lattice site index  ($r$).

Self-consistent Born solution is a saddle-point solution $Q$ of the action $S[Q,\omega]$. For simplicity, we will not explore general structures of the saddle-point solutions of specific tight-binding Hamiltonians $h_0$ in the chiral unitary class. Instead, we assume a (commonly used) {\it ansatz} $Q=\lambda \!\ 1_{NM\times NM}$. Real and imaginary parts of $\lambda$ physically mean an inverse of the mean free time and energy renormalization of diffusive particles, respectively. With periodic boundary conditions for $h_0$, the action in the momentum space may become
\begin{equation}
    S[\lambda 1_{N\times N},\omega]=-\frac{N M}{2\gamma^2}\lambda^* \lambda+ N \sum_{k} \!\ \Tr \ln
    \begin{pmatrix}
        \omega-i\lambda & -h_0(\bm{k}) \\
        -h^{*}_0(\bm{k}) & \omega-i\lambda^*
    \end{pmatrix}. \label{eq:scb-action}
\end{equation}
Here we assume that the eigenvalues of an $M$ by $M$ non-Hermitian matrix $h_0$ are given by complex values $h_0(\bm{k})$, where $k$ takes $M$ discrete values from the periodic boundary condition. The trace ($\operatorname{Tr}$) on the right-hand side is taken over the chiral index. The stationary point is determined by 
\begin{equation}
    0=\frac{\partial S}{\partial \lambda^*}=-\frac{N M}{2\gamma^2}\lambda+N\!\ \sum_{k} 
    \Tr \left[\begin{pmatrix}
        \omega-i\lambda & -h_0(\bm{k}) \\
        -h^{*}_0(\bm{k}) & \omega-i\lambda^*
    \end{pmatrix}^{-1}
    \begin{pmatrix}
        0 & 0 \\
        0 & -i
    \end{pmatrix}
    \right].
\end{equation}
The trace can be calculated as   
\begin{equation}
    \Tr \left[\begin{pmatrix}
        \omega-i\lambda & -h_0(\bm{k}) \\
        -h^*_0(\bm{k}) & \omega-i\lambda^*
    \end{pmatrix}^{-1}
    \begin{pmatrix}
        0 & 0 \\
        0 & -i
    \end{pmatrix}
    \right]=\sum_{\bm{k}}\frac{-i(\omega-i\lambda)}{(\omega-i\lambda)(\omega-i\lambda^*)-\epsilon_{k}^2},
\end{equation}
with $\epsilon^2_{k}=|h_{0}(\bm{k})|^2 \ge 0$. The saddle-point equation reduces to
\begin{equation}
    \lambda=-\frac{2i\gamma^2}{M}\sum_{k}\frac{\omega-i\lambda}{(\omega-i\lambda)(\omega-i\lambda^*)-\epsilon_{k}^2}.
\end{equation}
In the thermodynamic limit, we may replace the $k$-sum by an energy integral with a density of states  $\rho(\epsilon)  \equiv \frac{1}{M}\sum_k \delta(\epsilon-\epsilon_k)$,
\begin{equation}
    \lambda=-2i\gamma^2 \int_0^{\infty}d\epsilon\,\rho(\epsilon)\frac{\omega-i\lambda}{(\omega-i\lambda)(\omega-i\lambda^*)-\epsilon^2}. 
\end{equation}
Here, we set $\omega=0$, because the chiral symmetry is respected exclusively by zero-energy eigenstates of $H$, and the Anderson transition in chiral symmetry class depends solely on the zero-energy Green's function $G(\omega=0)$. The gap equation at $\omega=0$ suggests either $\lambda=0$ solution or $\lambda\ne 0$ solution that satisfies,
\begin{align}
1 = 2\gamma^2 \int^{\infty}_{0} d\epsilon \!\ \frac{\rho(\epsilon)}{|\lambda|^2+\epsilon^2}.  
\end{align}
When $\rho(\epsilon)\propto \epsilon^\alpha$ for  $\epsilon\simeq 0$ and $\alpha> 1$,  the gap equation for small $\gamma$  has only $\lambda=0$ solution (a solution of ballistic transport)~\cite{fradkin1986,syzranovHighDimensionalDisorderDrivenPhenomena2018}. Otherwise, the gap equation generally has a $\lambda\ne 0$ solution (a solution of diffusive transport). 

\subsection{Nonlinear sigma model}
     The action at $\omega=0$ has a continuous symmetry, and the $\lambda \ne 0$ solution has a degeneracy; 
\begin{equation}\label{eq:A14}
    \braket{{\cal Z}^N}_{\text{dis}}=\int \mathrm{D}[Q]\,\exp\left[-\frac{1}{2\gamma^2}\Tr \left(Q^\dagger Q\right)+\Tr \ln
    \begin{pmatrix}
        -iQ & -h_0 \\
        -h^\dagger_0 & -iQ^\dagger
    \end{pmatrix}
    \right]. 
\end{equation}
The action is invariant under the continuous transformation $Q\to U_1 Q U_2^\dagger$, with two $N$ by $N$ unitary matrices $U_1$ and $U_2$ that are independent of the spatial coordinate $r$. The symmetry dictates that the self-consistent Born (SCB) solution $Q=\lambda 1_{N\times N}$ must have degeneracy with $Q=\lambda U_1U_2^\dagger$. For $U_1=U_2$, they are identical. Thus, the degenerate SCB solutions form a manifold defined by a matrix group $\mathrm{U}(N)\times\mathrm{U}(N)/\mathrm{U}(N)\cong\mathrm{U}(N)$. The matrix group is called the Goldstone manifold. The group $\mathrm{U}(N)\times\mathrm{U}(N)$ in the numerator is called the symmetry group, while the group $\mathrm{U}(N)$ in the denominator is called the stabilizer group.

Since the SCB solution breaks the continuous symmetry, the solution is accompanied by gapless Goldstone modes -- diffusion mode. To describe the diffusion mode, let us take $Q=\lambda U$, where an $N$ by $N$ matrix $U$ is diagonal in $r$ but slowly varying in $r$. The partition function of the replicated fermion fields reduces to
\begin{align}
\braket{{\cal Z}^N}_{\text{dis}}
    &=\int \mathrm{D}[U]\,\exp\left[-\frac{|\lambda|^2 M N}{2\gamma^2} + \Tr \ln
    \begin{pmatrix}
        -i\lambda & -U^\dagger h_0 U \\
        -h^\dagger_0 & -i\lambda^*
    \end{pmatrix}\right].
\end{align}
On the right-hand side, the $r$-dependent $U$ does not commute with $h_0$, while the $r$-independent $U$ yields the SCB action, Eq.~(\ref{eq:scb-action}). To extract energy cost from the $r$-dependency in $U$, one can exercise a gradient expansion around the SCB Green's function $G^{-1}_0$, 
\begin{align}\label{eq:A17}
\braket{{\cal Z}^N}_{\text{dis}}
    &= \exp [-S_{\rm scb}] \int \mathrm{D}[U] \exp \left[\Tr \ln \left[1 + G_0 \begin{pmatrix} 
         0 & \Delta_{U} h_0 \\
         0 &  0
    \end{pmatrix}\right]\right] \nonumber \\
    & = \exp [-S_{\rm scb}] \int \mathrm{D}[U] \exp \left[ \Tr  \left[G_0 \begin{pmatrix} 
         0 & \Delta_{U} h_0 \\
         0 &  0
    \end{pmatrix}\right] - \frac{1}{2} \Tr  \left[\!\ G_0 \begin{pmatrix} 
         0 & \Delta_{U} h_0 \\
         0 &  0
    \end{pmatrix} G_0 \begin{pmatrix} 
         0 & \Delta_{U} h_0 \\
         0 &  0
    \end{pmatrix} \right] + \cdots \right] \nonumber \\
    & \equiv \exp [-S_{\rm scb}] \int \mathrm{D}[U] \exp [-S_{\rm NSLM}[U]],
\end{align}
with $S_{\rm scb} \equiv S [\lambda \!\ 1_{N\times N},\omega=0]$ and $\Delta_{U} h_0 \equiv -U^\dagger h_0 U + h_0$. The SCB Green's function $G_0$ at the zero energy ($\omega=0$) is given by,
\begin{align}
G^{-1}_0 = \begin{pmatrix} 
         -i\lambda &  - h_0\\
         -h^{\dagger}_0 &  -i\lambda^*
    \end{pmatrix}.
\end{align}

Since $\Delta_U h_0=0$ for the $r$-independent $U$, $\Delta_U h_0$ must reduce to a linear combination of $U^{-1}\nabla_{\mu} U$ in the continuum limit ($\mu=x,y,\cdots$). The form of $S_{\rm NLSM}[U]$ in terms of these gradient terms can be readily determined by a symmetry argument and local stability of the SCB solution. To this end, suppose that a disordered system considered here has (statistical) mirror symmetries with respect to spatial directions except for $z$, e.g., $(x,y,z) \rightarrow (-x,y,z)$ or $(x,-y,z)$. Possible scalar quantities one can construct out of $U^{-1}\nabla_{\mu} U$ and their products are $\Tr [U^{-1}\nabla_{\mu} U]$, $\Tr [U^{-1}\nabla_{\mu} U\cdot U^{-1}\nabla_{\nu} U]$,  $\Tr [U^{-1}\nabla_{\mu} U]\! \ \Tr [U^{-1}\nabla_{\nu} U]$,  and higher-order gradient terms~\cite{gadeReplicaLimitUnSOn1991,gadeAndersonLocalizationSublattice1993}. The mirror symmetries forbid cross terms ($\mu\ne \nu$) in  
$\Tr [U^{-1}\nabla_{\mu}U \cdot U^{-1}\nabla_{\nu}U]$ and $\Tr [U^{-1}\nabla_{\mu}U]\Tr [U^{-1}\nabla_{\nu}U]$. The symmetries also forbid $\Tr [U^{-1}\nabla_{\mu} U]$ with $\mu=x,y$. Thus, up to the second order in the expansion, the symmetry-allowed form of $S_{\rm NLSM}$ is given by 
\begin{align}
    S_{\rm NLSM}[U]=-\int\frac{\mathrm{d}^d r}{8\pi}\,\sum_{\mu}\left[\sigma_\mu\Tr \left(U^{-1}\nabla_\mu U\right)^2+c_\mu\Tr ^2\left(U^{-1}\nabla_\mu U\right)-\chi \!\  \Tr \left(U^{-1}\nabla_z U\right)\right].
\end{align}
Notably, $\sigma_{\mu}$ and $c_{\mu}$ must be positive and $\chi$ must be real; otherwise, the spatially uniform saddle-point solution $Q=\lambda 1_{N\times N}$ cannot be a local energy minimum of the action $S[Q,\omega]$. In the absence of any other statistical symmetries that change the sign of $z$, e.g. $(x,y,z)\rightarrow (x,y,-z)$, real-valued $\chi$ is generally finite. 

\subsection{Nodal-line Dirac semimetal model}
     Chiral symmetric models with the 1D weak topological term can be exemplified by a disordered nodal-line semi-metal defined on the cubic lattice,~\cite{Ryu2010,schnyder2011,fulga2012a,mondragon-shem2014,luo2020,claes2021,xiaoAnisotropicTopologicalAnderson2023,silva2025}.  
\begin{align}
    H &=\sum_{\bm{r}}\bigg[\epsilon^{\prime}_{\bm r} c^{\dagger}_{\bm r} 
    \sigma_x c_{\bm r} + \epsilon^{\prime\prime}_{\bm r}c^{\dagger}_{\bm r} 
    \sigma_y c_{\bm r} + \Delta c^\dagger_{\bm{r}}\sigma_xc_{\bm{r}}+ \nonumber \\
    &  \ \ \ \  + \left(t_\perp c^\dagger_{\bm{r}+\hat{\bm{x}}}\sigma_xc_{\bm{r}}+t_\perp c^\dagger_{\bm{r}+\hat{\bm{y}}}\sigma_xc_{\bm{r}}+it_\parallel c^\dagger_{\bm{r}+\hat{\bm{z}}}\sigma_y c_{\bm{r}}+t'_{\parallel}c^\dagger_{\bm{r}+\hat{\bm{z}}}\sigma_xc_{\bm{r}}+\text{h.c.}\right)\bigg]. \label{nlsm-1}
\end{align}
Here ${\bm r}=(x,y,z)$ is the 3D coordinate of the cubic lattice site, and $\sigma_{x}$ and $\sigma_y$ are the 2-by-2 Pauli matrices in the chiral space. $\epsilon^{\prime}_{\bm r}$ and $\epsilon^{\prime\prime}_{\bm r}$ are random real numbers representing the disorder. As in the previous section, we assume the Gaussian distribution of $V_{\bm r}\equiv\epsilon^{\prime}_{\bm r} + i\epsilon^{\prime\prime}_{\bm r}$: $\langle V^*_{\bm r} V_{{\bm r}^{\prime}}\rangle = \gamma^2 \delta_{\bm r,\bm r^{\prime}}$. The Hamiltonian has only chiral symmetry,  belonging to the chiral unitary class. With $H \equiv h \sigma_{+} + h^{\dagger}\sigma_{-}$, the off-diagonal non-Hermitian tight-binding model $h$ has a non-reciprocal (Hatano-Nelson) hopping along the $z$ direction, $|t_{\parallel}+t^{\prime}_{\parallel}|\ne |t_{\parallel}-t^{\prime}_{\parallel}|$, while tight-binding hopping within the $xy$ plane is reciprocal,
\begin{align}
h &= \sum_{\bm r} \Big[V_{\bm r} a^{\dagger}_{\bm r} a_{\bm r} + \Delta a^\dagger_{\bm{r}} a_{\bm{r}}+ t_\perp \sum_{\mu=x,y}( a^\dagger_{\bm{r}+\hat{\bm{\mu}}}a_{\bm{r}}+a^\dagger_{\bm{r}-\hat{\bm{\mu}}}a_{\bm{r}}) + (t_\parallel + t^{\prime}_{\parallel}) a^\dagger_{\bm{r}+\hat{\bm{z}}} a_{\bm{r}}+(-t_\parallel + t^{\prime}_{\parallel}) a^\dagger_{\bm{r}-\hat{\bm{z}}} a_{\bm{r}} \Big]. \label{nlsm-2}
\end{align}
 In the clean limit $(V_{\bm r}=0)$, the Fourier transform of $h_0$ is given by a complex number,
 \begin{align}
 h_0({\bm k}) \equiv \Delta + 2t_{\perp} (\cos k_x + \cos k_y) + 2i t_{\parallel} \sin k_z 
 + 2t^{\prime}_{\parallel} \cos k_z. 
 \end{align}
Thus, $H_0\equiv  h_0 \sigma_{+} + h^{\dagger}_0\sigma_{-}$ has an energy gap around zero for (i) $|\Delta-4t_{\perp}|<2|t^{\prime}_{\parallel}|$ and $|\Delta+4t_{\perp}|<2|t^{\prime}_{\parallel}|$, (ii) $\Delta-4t_{\perp}<-2|t^{\prime}_{\parallel}|$ and $\Delta+4t_{\perp}<-2|t^{\prime}_{\parallel}|$, and (iii) $\Delta-4t_{\perp}>2|t^{\prime}_{\parallel}|$ and $\Delta+4t_{\perp}>2|t^{\prime}_{\parallel}|$. For the other tight-binding parameters,  $H_0$ has zero-energy eigenstates whose crystal momenta ${\bm k}=(k_x,k_y,k_z)$ shape a closed line on the $k_z=0$ or $k_z=\pi$ planes. Around the line of the zero-energy states, the low-energy states has a linear energy-momentum dispersion --- nodal-line Dirac semimetal. 

The 1D weak topological indices ${\bm \chi}\equiv (\chi_x,\chi_y,\chi_z) $ of $H=h \sigma_{+} + h^{\dagger}\sigma_{-}$ in chiral symmetry classes are most generally characterized by the Lyapunov exponents (LEs) of the off-diagonal non-Hermitian Hamiltonian $h$ \cite{luca2003,fulga2012a,xiaoAnisotropicTopologicalAnderson2023,zhaoTopologicalEffectAnderson2024c}. The 1D weak topological index along the $\mu$ direction is given by a ratio between a number $N_{+,\mu}$ of positive LEs of $h$ along $\mu$ and a number $N_{-,\mu}$ of negative LEs of $h$ along $\mu$: $\chi_{\mu}/8\pi = (N_{+,\mu}-N_{-,\mu})/(N_{+,\mu}+N_{-,\mu})$. Due to the non-reciprocal hopping along $z$,  the LEs of Eq.~(\ref{nlsm-2}) along $z$ have an asymmetric distribution around zero: $\chi_{z}$ can be non-zero for larger non-reciprocal hopping along $z$~\cite{xiaoAnisotropicTopologicalAnderson2023}. Meanwhile, the 1D weak topological indices of Eq.~(\ref{nlsm-2}) along the $x$ and $y$ directions are strictly zero because of the statistical symmetry: $\chi_{x}=\chi_y=0$~\cite{xiaoAnisotropicTopologicalAnderson2023}. Namely, an ensemble of $h$ is symmetric under a transposition followed by a mirror $M_{xy}$ with respect to the $xy$ plane, 
\begin{align}
\Big\{ \!\ M^{\dagger}_{xy} h^{T} M_{xy}  \!\ \Big| \langle V_{\bm r}\rangle =0, \langle V^*_r V_{r^{\prime}}\rangle = \gamma^2 \delta_{r,r^{\prime}} \Big\} = \Big\{\!\ h \!\ \Big| \langle V_{\bm r}\rangle =0, \langle V^*_r V_{r^{\prime}}\rangle = \gamma^2 \delta_{r,r^{\prime}} \Big\},  
\end{align}
with $(M_{xy})_{\bm r,\bm r^{\prime}} = \delta_{x,x^{\prime}}\delta_{y,y^{\prime}} \delta_{z,-z^{\prime}}$. Since the transposition changes the sign of the LEs along all directions, while the mirror changes the sign of the LEs only along $z$, the statistical symmetry requires the LEs along $x$ and $y$  to be symmetric around zero. This leads to $N_{+,\mu}=N_{-,\mu}$ for $\mu=x,y$ and $\chi_x=\chi_y=0$.  

The non-linear sigma model with $\chi_z\ne 0$ can be readily derived from the nodal line semimetal model by the use of $\bm{k}\cdot\bm{p}$ expansion, self-consistent Born approximation, and gradient expansion. To this end, let us consider a specific nodal-line Dirac semimetal region; $\Delta< -2t^{\prime}_{\parallel}\lesssim\Delta+4t_{\perp}<2t^{\prime}_{\parallel}$ with positive $t^{\prime}_{\parallel}$ and $t_{\perp}$. Thereby, the crystal momenta ${\bm k}$ of the zero-energy eigenstates of $H_0$ form a small ring at $\Delta + 2t_{\perp}(\cos k_x+\cos k_y)+2t^{\prime}_{\parallel} =0$ and $k_z=0$, encircling the $\Gamma$ point. In such a semimetal region, low-energy disordered states can be legitimately analyzed using the following effective continuum model, 
\begin{align}
H_{\rm eff} &= \int d^3 {\bm r} \!\ \!\  \psi^{\dagger}({\bm r}) \bigg[ \epsilon^{\prime}_{\bm r} 
\sigma_x + \epsilon^{\prime\prime}_{\bm r} \sigma_y + M \sigma_x 
+ t_{\perp} \Big( \partial^2_x + \partial^2_y\Big) + 2t_{\parallel} i \partial_z \sigma_y 
 \bigg] \psi({\bm r}) \nonumber \\
& = \int d^3 {\bm r} \!\ \!\  \psi^{\dagger}({\bm r}) \bigg[ 
\Big(\hat{h}^{\rm eff}_0\big({\bm \nabla}_{\bm r}\big) + V_{\bm r} \Big) \sigma_{+} + \Big(\hat{h}^{\rm eff \!\ \!\  \dagger}_0\big({\bm \nabla}_{\bm r}\big) + V^*_{\bm r} \Big) \sigma_{+}\bigg] \psi({\bm r}), \label{nlsm-3}
\end{align}
with 
\begin{align}
\hat{h}^{\rm eff}_{0}\big({\bm \nabla}_{\bm r}\big) = M + t_{\perp} (\partial^2_x + \partial^2_y) + 2t_{\parallel} \partial_z , 
\end{align}
and $M  \equiv \Delta + 4t_{\perp}+2t^{\prime}_{\parallel} \gtrsim 0$.  Here, we expand $h_{0}({\bm k})$ around ${\bm k}=0$ up to the lowest order in each $k_{\mu}$, and replace the small ${\bm k}=(k_x,k_y,k_z)$ with the spatial gradient $-i{\bm \nabla}_{\bm r}=-i(\partial_x,\partial_y,\partial_z)$ respectively. Since, for the small positive $M$, the DOS near the zero energy vanishes linearly with energy $\rho_0(\epsilon) \propto \epsilon$, small disorder allows the SCB solution $Q=\lambda 1_{N\times N}$ with non-zero $\lambda$. Thus, by regarding $Q(\bm r)= \lambda U({\bm r})$ as a slowly varying unitary matrix, we exercise the gradient expansion in a disorder-averaged partition function for the effective continuum model, 
\begin{equation}
    \braket{\mathcal{Z}^N}_{\text{dis}}=\int\mathrm{d}[Q]\,\exp\left\{-\frac{1}{2\gamma^2}\int\mathrm{d}^3r\Tr\left[Q^\dagger(\bm{r})Q(\bm{r})\right]+\int\mathrm{d}^3r\,\Tr\ln
    \begin{pmatrix}
        -iQ(\bm{r}) & -\hat{h}^{\rm eff}_0\big({\bm \nabla}_{\bm r}\big) \\
        -\hat{h}^{{\rm eff}\!\ \!\ \dagger}_0\big({\bm \nabla}_{\bm r}\big) & -iQ^\dagger(\bm{r})
    \end{pmatrix}
    \right\}.
\end{equation}
This leads to a low-energy effective action for gapless diffusion modes, as follows: 
\begin{align}\label{eq:A25}
    S_{\text{NLSM}}[U]=& - \int \mathrm{d}^3 r \int  \mathrm{d}^3r^{\prime}\, \Tr\Bigg[G^{\rm eff}_0({\bm r}-{\bm r}^{\prime})
    \begin{pmatrix}
        0 & \Delta_{U({\bm r}^{\prime})} \hat{h}^{\rm eff}_0\big({\bm \nabla}_{{\bm r}^{\prime}}\big)   \\
        0 & 0
    \end{pmatrix} \delta^3({\bm r}^{\prime}-{\bm r})
    \Bigg]   \nonumber \\ 
    & +\frac{1}{2}\int \mathrm{d}^3 r_1 \int \mathrm{d}^3 r^{\prime}_1  \int \mathrm{d}^3 r_2 \int \mathrm{d}^3 r^{\prime}_2\,  \Tr\Bigg[G^{\rm eff}_0({\bm r}_1-{\bm r}^{\prime}_1)
    \begin{pmatrix}
        0 & \Delta_{U({\bm r}^{\prime}_1)} \hat{h}^{\rm eff}_0\big({\bm \nabla}_{{\bm r}^{\prime}_1}\big) \\
        0 & 0
    \end{pmatrix}  \nonumber \\
    & \hspace{2cm} \delta^3({\bm r}^{\prime}_1-{\bm r}_2) G^{\rm eff}_{0}({\bm r}_2-{\bm r}^{\prime}_2)
    \begin{pmatrix}
        0 & \Delta_{U({\bm r}^{\prime}_2)} \hat{h}^{\rm eff}_0\big({\bm \nabla}_{{\bm r}^{\prime}_2}\big)  \\
        0 & 0
    \end{pmatrix} \delta^3({\bm r}^{\prime}_2-{\bm r}_1) 
    \Bigg]+\cdots,
\end{align}
where $G^{\rm eff}_0({\bm r})$ is the self-consistent Born Green's function,
\begin{align}
\left(\begin{array}{cc}
-i\lambda & -\hat{h}^{\rm eff}_{0}\big({\bm \nabla}_{\bm r}\big) \\
-\hat{h}^{{\rm eff}}_0\big({\bm \nabla}_{\bm r}\big)^{\dagger} & -i\lambda \\ 
\end{array}\right) G^{\rm eff}_0({\bm r}) = \delta^3({\bm r})  \left(\begin{array}{cc}
1 & 0 \\
0 & 1 \\
\end{array}\right). \label{green}
\end{align}
$\Delta_{U(\bm{r})} \hat{h}^{\rm eff}_0({\bm \nabla}_{\bm r})$ is a function of  ${\bm r}$ and ${\bm \nabla}_{\bm r}$,  
\begin{align}
\Delta_{U(\bm{r})} \hat{h}^{\rm eff}_0 \big({\bm \nabla}_{\bm r}\big)
= - U^{\dagger}({\bm r})\, \hat{h}^{\rm eff}_0({\bm \nabla}_{\bm r})\, U({\bm r}) + \hat{h}^{\rm eff}_0.  
\end{align}
${\rm Tr}$ is taken over the chiral and replica indices.

    The 1D weak topological term in the effective action comes from the first-order gradient term, 
\begin{align}
    &-\int\mathrm{d}^3r\int \mathrm{d}^3r^{\prime}\,\Tr\left[
    G^{\rm eff}_{0}({\bm r}-{\bm r}^{\prime})
    \begin{pmatrix}
        0 &  \Delta_{U(\bm{r}^{\prime})} \hat{h}^{\rm eff}_0 \big({\bm \nabla}_{{\bm r}^{\prime}}\big)\\
        0 & 0
    \end{pmatrix}\delta^3({\bm r}^{\prime}-{\bm r})\right] \nonumber \\
    & \ \ =\int\mathrm{d}^3r^{\prime}\,\Tr\left[
    G^{\rm eff}_0({\bm 0})
    \begin{pmatrix}
        0 & 2t_\parallel U(\bm{r}^{\prime})^\dagger\partial_{z^{\prime}} U(\bm{r}^{\prime}) \\
        0 & 0
    \end{pmatrix}\right]+\cdots, \nonumber \\
    & \ \ \equiv  \frac{\chi_z}{8\pi} \int\mathrm{d}^3r\, \Tr\left[U(\bm{r})^\dagger\partial_z U(\bm{r})\right] + \cdots. 
\end{align}
Here, higher-order spatial derivative terms are omitted as $\cdots$. The trace in the second line is over the chiral and replica indices, while the trace in the third line is only over the replica index. Noting that $G^{\rm eff}_0({\bm r}-{\bm r}^{\prime}={\bm 0})$ is given by ${\bm k}$-integral of the Fourier transform $G^{\rm eff}_0({\bm k})$ of the Green's function, one can write the 1D weak topological coefficient in terms of the trace only over the chiral index,
\begin{align}
    \frac{\chi_z}{8\pi} &= \int\frac{\mathrm{d}^3 k}{(2\pi)^3}\,\Tr\left[
    G^{\rm eff}_0({\bm k})
    \begin{pmatrix}
        0 & 2t_\parallel  \\
        0 & 0
    \end{pmatrix}\right]  \nonumber \\
    &= i\int\frac{\mathrm{d}^3 k}{(2\pi)^3}\,\Tr\left[G^{\rm eff}_0(\bm{k})\, \partial_{k_z}G^{{\rm eff},-1}_0(\bm{k})\left(\begin{array}{cc}
   0 & 0 \\
   0 & 1 \\
   \end{array}\right)
    \right]. \label{1Dwinding}
\end{align} 
In the second line, we use an inverse of $G^{\rm eff}_0({\bm k})$ from Eq.~(\ref{green}), 
\begin{align}
G^{\rm eff,-1}_0({\bm k}) = \left(\begin{array}{cc} 
-i\lambda &- h^{\rm eff}_{0}({\bm k}) \\
-h^{\rm eff}_0({\bm k})^* & -i\lambda \\
\end{array}\right), 
\end{align}
with 
\begin{align}
h^{\rm eff}_0({\bm k}) = M - t_{\perp} (k^2_x + k^2_y) + 2t_{\parallel} ik_z. 
\end{align}
Eq.~(\ref{1Dwinding}) can be regarded as the 1D weak topological index evaluated within the self-consistent Born approximation.  In fact, when $\lambda=0$, it reduces to an expression consistent with the 1D weak topological index in the clean limit~\cite{Ryu2010,schnyder2011,mondragon-shem2014,luo2020,claes2021,xiaoAnisotropicTopologicalAnderson2023}, i.e. 
\begin{align}
\lim_{\lambda\rightarrow 0}\frac{\chi_z}{8\pi} = i \int \frac{\mathrm{d}^3k}{(2\pi)^3} \partial_{k_z} \Big({\rm Log}\, \big[h^{\rm eff}_0({\bm k})\big]\Big) = \frac{S_{k_xk_y}}{4\pi^2}. 
\end{align}
The right hand side is given by an area $S_{k_xk_y}$ inside the zero-energy ring on the $k_z=0$ plane;
\begin{align}
S_{k_xk_y} = \int_{\Delta + 2t_{\perp}(\cos k_x+\cos k_y)+2t^{\prime}_{\parallel} >0} \mathrm{d}k_x\mathrm{d}k_y.  
\end{align}
The expression shows that $\chi_z$ is finite if the renormalized zero-energy ring has a finite area $S_{k_x k_y}$ inside the ring in momentum space. Since the nodal ring can be annihilated only by shrinking it into a point node, the expression indicates that $\chi_z \ne 0$ is a robust property of the chiral-symmetric NLD semimetals.

\section{Saddle-point equation of the nonlinear sigma model}\label{app:saddle_point}
In this Appendix, we derive a set of saddle-point equations of the chiral NLSM, Eqs~(\ref{eq:action_with_Ansatz},\ref{eq:saddle-point_equation1},\ref{eq:saddle-point-equation2}), as well as Eq.~(\ref{eq:field-strength}). We also discuss a general feature of $P(\bm{r}) \equiv |p(\bm{r})\rangle \langle p(\bm{r})|$ that satisfies the saddle-point equations, and how the Poisson equation for $\phi(\bm{r})$ can be obtained from them. The derivation can be applied to all three chiral symmetry classes. $Q(\bm{r})$ in the chiral classes has unit complex numbers $e^{i\phi_n(\bm{r})}$ as its eigenvalues, 
\begin{equation}
    Q(\bm{r})=\sum_{n=1}^{N}e^{i\phi_n(\bm{r})}\ket{p_n(\bm{r})}\bra{p_n(\bm{r})}. \nonumber 
\end{equation}
$|p_n(\bm{r})\rangle$ is a corresponding eigenvector. General saddle-point solutions of the NLSM could be obtained from its derivatives with respect to all eigenvalues and eigenvectors of the $Q$ field. In saddle-point solutions thus obtained, each eigenvalue $e^{i\phi_n(\bm{r})}$ could have multiple vortex excitations separately. When such vortex excitations have no spatial overlap with one another, however, one may utilize an $r$-dependent basis change in such a way that only one eigenvalue $e^{i\phi(\bm{r})}$ has all the vortex excitations, while the others do not. To introduce such a generic $Q$ configuration as the saddle-point solution, we may use the following ansatz; 
\begin{align}
    Q(\bm{r})&=e^{i\psi(\bm{r})}+\left[e^{i\phi(\bm{r})}-e^{i\psi(\bm{r})}\right]|p(\bm{r})\rangle\langle p(\bm{r})| \nonumber \\
             &= e^{i\psi(\bm{r})} \left(1+ \left[e^{i\phi(\bm{r})-i\psi(\bm{r})}-1\right]|p(\bm{r})\rangle\langle p(\bm{r})|\right).
    \label{Eq:B-1}
\end{align}
In the ansatz, only one eigenvalue $e^{i\phi({\bm r})}$, which an eigenstate $|p(\bm{r})\rangle$ belongs to, has vortex excitations, while the other eigenvalues $e^{i\psi(\bm{r})}$ are degenerate, and do not have any vortices.  Since the overall factor $e^{i\psi(\bm{r})}$ in the right hand side can be absorbed into the spin-wave part of the $Q$ field, we may also use Eq.~(\ref{eq:Ansatz}) by replacing  $e^{i\phi(\bm{r})-i\psi(\bm{r})}$ by $e^{i\phi(\bm{r})}$.  From Eq.~(\ref{eq:Ansatz}) with $P(\bm{r}) \equiv \ket{p(\bm{r})}\bra{p(\bm{r})}$, we obtain
\begin{align}
    \nabla_\mu Q &=i\nabla_\mu \phi\, \  e^{i\phi}P+\left(e^{i\phi}-1\right)\nabla_\mu P, \nonumber \\
    Q^{-1}\nabla_\mu Q&=i\nabla_\mu \phi P+2(1-\cos\phi)P\nabla_\mu P+(e^{i\phi}-1)\nabla_\mu P. \label{Eq:B-1a}
\end{align}
We define $u\equiv \Tr P$, which is $1$ for class AIII and CII, and $2$ for class BDI. With the use of $ \Tr (P)=u$,  $\Tr (\nabla_\mu P)=\nabla_\mu \Tr (P)=0$, $\Tr (P\nabla_\mu P)=\Tr (\nabla_\mu P P)=\frac{1}{2}\Tr (\nabla_\mu P)=0$, $P^2=P$, $\nabla_\mu PP+P\nabla_\mu P=\nabla_\mu P$, and 
\begin{align}
    &\Tr (P\nabla_\mu P)^2=\Tr (\nabla_\mu P P\nabla_\mu P)-\Tr (\nabla_\mu P P^2\nabla_\mu P)=0,  \nonumber \\
    &\Tr (\nabla_{\mu} P)^2 = 2 \Tr (P\nabla_{\mu} P \nabla_{\mu} P), \nonumber 
\end{align}
we obtain the following formulas, 
\begin{align}
    &\Tr \left(Q^{-1}\nabla_\mu Q\right)=i u \nabla_\mu\phi,  \nonumber \\
    &\Tr \left(Q^{-1}\nabla_\mu Q\right)^2=-u\left[(\nabla_\mu \phi)^2+2(1-\cos\phi)\Tr (\nabla_{\mu} P)^2\right]. 
    \nonumber 
\end{align}
Thus, the action with the saddle-point ansatz, Eq.~(\ref{eq:Ansatz}), is given by
\begin{equation}
    S=\frac{u}{8\pi s}\int\mathrm{d}^3r\,\sum_{\mu}\left[(\sigma_\mu+u c_\mu)(\nabla_\mu \phi)^2+2\sigma_\mu(1-\cos\phi)\Tr (\nabla_\mu P)^2+i\chi_\mu\nabla_\mu \phi\right]. \label{Eq:B0}
\end{equation}
Derivatives of Eq.~(\ref{Eq:B0}) with respect to the U(1) phase $\phi(\bm{r})$ and matrix elements of $P(\bm{r})$ give 
\begin{equation}
    \sum_{\mu}(\sigma_{\mu}+u c_{\mu})\nabla^2_{\mu}\phi-\sum_{\mu}\sigma_{\mu} \sin\phi \Tr (\nabla_{\mu} P)^2=0, \label{Eq:B1}
\end{equation}
and 
\begin{equation}
   \sum_{\mu} \sigma_{\mu}(1-\cos\phi)\nabla^2_{\mu} P+\sum_{\mu}\sigma_{\mu} \sin\phi \nabla_{\mu}\phi\nabla_{\mu} P=0, \label{Eq:B2}
\end{equation}
respectively. By multiplying \cref{Eq:B2} by $P$ and taking the trace, we can also have
\begin{equation}
   \sum_{\mu} \sigma_{\mu}(1-\cos\phi) \!\ \Tr (P\nabla^2_{\mu} P)+ \sum_{\mu} \sigma_{\mu}\sin\phi \!\ \nabla_{\mu}\phi \!\ \Tr (P\nabla_{\mu} P)=0. \label{Eq:B3}
\end{equation}
As $\Tr (P\nabla_{\mu} P)=0$, and $\Tr (P\nabla^2_{\mu} P)=\nabla_{\mu}\Tr (P\nabla_{\mu} P)-\Tr (\nabla_{\mu} P)^2=-\Tr (\nabla_{\mu} P)^2$, Eq.~(\ref{Eq:B3}) means either $\sum_{\mu}\sigma_{\mu}\Tr (\nabla_{\mu} P)^2=0$ or $\phi=2\pi \mathbb{Z}$. For $\phi=2\pi \mathbb{Z}$,  $Q$ in Eq.~(\ref{eq:Ansatz}) becomes independent of $P(\bm{r})$, where we can choose $P(\bm{r})$ freely. Note that $P(\bm{r})$ that satisfies $\sum_{\mu}\sigma_{\mu}\Tr (\nabla_{\mu} P)^2=0$ becomes independent of $r$; $P(\bm{r})=P_0$. This is because as $\sigma_{\mu}\ge 0$, $\sum_{\mu}\sigma_{\mu}\Tr (\nabla_{\mu} P)^2=\sum_{\mu}\sum^N_{i,j=1} \sigma_{\mu} |\nabla_{\mu} P_{ij}({\bm r})|^2=0$ requires that $P(\bm{r})$ is independent of $\bm{r}$. For $\phi=2\pi \mathbb{Z}$, we choose $P(\bm{r})=P_0$, so that $\sum_{\mu}\sigma_{\mu}\Tr (\nabla_{\mu} P)^2=0$ is satisfied entirely. Then, Eq.~\eqref{Eq:B1} reduces to a Poisson equation, 
\begin{equation}
    (\sigma_{\mu}+ u c_{\mu})\nabla^2_{\mu}\phi=0.
\end{equation}
 Note also that when $\phi$ is indefinite such as in vortex cores, Eqs.~(\ref{Eq:B2},\ref{Eq:B3}) do not have to be satisfied, where $P(\bm{r})$ is not specified: $P(\bm{r})$ can depend on $r$ along vortex loops. From Eq.~(\ref{Eq:B-1a}), the field strength $F_{\mu}\equiv \epsilon_{\mu\nu\lambda}\nabla_{\nu}h_{\lambda}+ i\epsilon_{\mu\nu\lambda}h_{\nu}h_{\lambda}$ with  $h_{\lambda} \equiv -iQ^{-1}\nabla_{\lambda} Q$ can be calculated,
\begin{align}
F_{\mu}(\bm{r}) \equiv \epsilon_{\mu\nu\lambda} 
\nabla_{\nu}\nabla_{\mu} \phi(\bm{r}) \!\ P(\bm{r}).  
\end{align} 

\section{Variational study of $\mathrm{U}(N)$ type-II superconductor model}\label{app:variational_method}
In this Appendix, we elaborate the variational analysis of the superconducting (Meissner) phase in the $\mathrm{U}(N)$ type-II superconductor model for $\chi=0$ case, Eqs.~(\ref{eq:dual_discretized},\ref{eq:dual_discretized_u1},\ref{eq:dual_discretized_suN},\ref{eq:dual_discretized_Sy}). In the superconducting phase, the $\psi$ field assumes a spatially uniform value, and the spin-wave fluctuation around the uniform configuration can be absorbed into the U(1) gauge field. The action thus obtained takes the following form in the continuum limit, 
\begin{equation}
    S=2\pi\int \mathrm{d}^3r\,\left[\frac{\left(\bm{\nabla}\times\bm{\theta}^0
    %-\frac{\sqrt{N}}{8\pi}\chi
    \right)^2}{\sigma+Nc}+\sum_{a}\frac{\left(\bm{\nabla}\times\bm{\theta}^a\right)^2}{\sigma}\right]-2t \Lambda^{3}\sum_{\mu}\int \mathrm{d}^3 r \int \mathrm{d}|p(\bm{r})\rangle \cos\Tr (2\pi\Lambda^{-1}\Theta_\mu P).
\end{equation}
Thanks to the $|p(\bm{r})\rangle$ integral, the U(1) and $\mathrm{SU}(N)$ gauge fields are decoupled from one another at the quadratic level in the Taylor expansion of the $\cos$ term. Thus, the Meissner phase can be described by the following mean-field action with finite $m$,
\begin{align}
\overline{S} =  2\pi\int \mathrm{d}^3r\left\{\left[\frac{\left(\bm{\nabla}\times\bm{\theta}^0\right)^2}{\sigma+Nc}+m^2\left(\bm{\theta}^0\right)^2\right]+\sum_{a=1}^{N^2-1}\left[\frac{\left(\bm{\nabla}\times\bm{\theta}^a\right)^2}{\sigma}+m^2\left(\bm{\theta}^a\right)^2\right]\right\}.
\end{align}
Here we assume the same mass $m$ for the U(1) and S$\mathrm{U}(N)$ gauge field (see the replica limit of Eq.~(\ref{eq:action_for_SC})). To obtain a gap equation for $m$, we exercise a variational principle which states that a trial free energy $F(m)$ defined below must be always greater than the true free energy $F_{\rm true}\equiv -\ln Z \equiv \int \mathrm{D}[\Theta] \exp\left[-S\right]$,    
\begin{align}
&F(m) \equiv  -\ln \overline{Z} + \langle S - \overline{S}\rangle_{\overline{S}}, 
\end{align}
and 
\begin{align}
\langle \cdots \rangle_{\overline{S}} &\equiv \frac{1}{\overline{Z}} \int \mathrm{D}[\Theta] 
\cdots \exp \left[-\overline{S}\right], \nonumber \\
\overline{Z} &\equiv \int \mathrm{D}[\Theta] 
 \exp \left[-\overline{S}\right].  
\end{align}
The variational principle (Bogoliubov inequality) is rooted in the convex nature of the exponential function, which implies that an optimal value of $m$ can be obtained by minimizing $F(m)$ with respect to $m$\cite{chaikin_lubensky1995,nagaosa1999,herbutModernApproachCritical2007}. 

Notably, $F(m)$ reaches zero in the replica limit[see below], and so does $F_{\rm true}$ by the inequality. Since $F_{\rm true}=-\ln \langle {\cal Z}^N\rangle_{\rm dis}$ in Eq.~(\ref{eq:replica}), $\lim_{N\rightarrow 0} F_{\rm true}=0$ allows us to express the disorder-averaged free energy by $-\langle \ln {\cal Z }\rangle_{\rm dis} = \lim_{N\rightarrow 0}F_{\rm true}/N$.  Correspondingly, we regard $\lim_{N\rightarrow 0}F(m)/N$ as a trial function of the disorder-averaged free energy and minimize this trial function with respect to $m$. The following Green's functions, as well as their trace and determinant, facilitate an evaluation of $F(m)$,
\begin{align}
    G_{\mu\nu}^a(\bm{r}-\bm{r}') 
    &\equiv \int\frac{\mathrm{d}^3k}{(2\pi)^3}\,\langle \tilde{\theta}^a_\mu(-\bm{k})\tilde{\theta}^a_\nu(\bm{k})\rangle_{\overline{S}} \!\ \!\ e^{i\bm{k}\cdot(\bm{r}-\bm{r}')}\nonumber \\
    &\equiv\int\frac{\mathrm{d}^3k}{(2\pi)^3}\,\tilde{G}_{\mu\nu}^a(\bm{k}) \!\ \!\ e^{i\bm{k}\cdot(\bm{r}-\bm{r}')},
\end{align}
with $a=0,1,\cdots, N^2-1$. The Fourier transforms $\tilde{G}^a_{\mu\nu}({\bm k})$ of the Green's function are given by   
\begin{equation}
    \tilde{G}^0_{\mu\nu}(\bm{k})=\frac{1}{4\pi}\frac{1}{k^2+(\sigma+Nc)m^2}\left[(\sigma+Nc)\delta_{\mu\nu}+\frac{k_\mu k_\nu}{m^2}\right],
\end{equation}
\begin{equation}
    \tilde{G}^{a\ne 0}_{\mu\nu}(\bm{k})=\frac{1}{4\pi}\frac{1}{k^2+\sigma m^2}\left(\sigma\delta_{\mu\nu}+\frac{k_\mu k_\nu}{m^2}\right). 
\end{equation}
The trace and determinant of these 3 by 3 matrices are as follows, e.g.  
\begin{align}
        \Tr  \tilde{G}^{a\ne 0}(\bm{k})&=\frac{1}{4\pi}\frac{k^2+3\sigma m^2}{m^2(k^2+\sigma m^2)}, \nonumber \\
        \det\tilde{G}^{a\ne 0}(\bm{k}) &=\frac{\sigma^2}{64\pi^3}\frac{1}{m^2(k^2+\sigma m^2)^2}.
\end{align}
    
We calculate the density of the trial free energy $F(m)$ in the replica limit, 
\begin{equation}
    f(m)=\lim_{N\to 0}\frac{F(m)}{NV} = -\lim_{N\rightarrow 0}\frac{\ln \overline{Z}}{NV} + \lim_{N\rightarrow 0}\frac{\langle S-\overline{S}\rangle_{\overline{S}}}{NV}
\end{equation}
where the $m$-dependent part of $\ln \overline{Z}/V$ is given by the determinant of the Green's function, 
\begin{align}
    & \frac{1}{V}\ln \overline{Z}=\frac{1}{2}\sum_{a=0}^{N^2-1}\int\frac{ \mathrm{d}^3k}{(2\pi)^3}\,\ln\det \tilde{G}^a(\bm{k}) \nonumber \\
    & =-\int\frac{ \mathrm{d}^3 k}{(2\pi)^3}\,\ln\frac{k^2+(\sigma+Nc)m^2}{k^2+\sigma m^2}-N^2\int\frac{\mathrm{d}^3 k}{(2\pi)^3}\left[\ln m+\ln (k^2+\sigma m^2)\right],
\end{align}
In the replica limit, it becomes
\begin{align}
    \lim_{N\to 0}\frac{1}{NV}\ln \overline{Z} &=-\lim_{N\to 0}\frac{2}{N}\int\frac{\mathrm{d}^3 k}{(2\pi)^3}\,\ln\frac{k^2+(\sigma+Nc)m^2}{k^2+\sigma m^2} \nonumber \\
& =-\int\frac{\mathrm{d}^3 k}{(2\pi)^3}\,\frac{2cm^2}{k^2+\sigma m^2}.
\end{align}

$\langle S-\overline{S}\rangle_{\overline{S}}$ is given by 
\begin{align}
    \langle S-S_0 \rangle_{\overline{S}} =-2t\Lambda^3\sum_{\mu}\int \mathrm{d}^3r \int \mathrm{d}|p(\bm{r})\rangle \langle \!\ \cos\Tr (2\pi\Lambda^{-1}\Theta_\mu P)\rangle_{\overline{S}}-2\pi m^2 \sum^{N^2-1}_{a=0} \sum_{\mu=x,y,z}
    \int \mathrm{d}^3 r \!\ \langle \theta^a_{\mu}(\bm{r}) \theta^a_{\mu}(\bm{r})\rangle_{\overline{S}}, \label{eq:20}
\end{align}
where
\begin{equation}
    \sum_{\mu}\langle \theta^0_{\mu}(\bm{r})\theta^0_{\mu}(\bm{r})\rangle_{\overline{S}}=\int\frac{\mathrm{d}^3k}{(2\pi)^3}
    \Tr \tilde{G}^0(\bm{k})=\frac{1}{4\pi}\int\frac{\mathrm{d}^3k}{(2\pi)^3}\frac{k^2+3(\sigma+Nc) m^2}{m^2[k^2+(\sigma+Nc) m^2]},
\end{equation}
\begin{equation}
     \sum_{\mu}\langle \theta^{a\ne 0}_{\mu}(\bm{r})\theta^{a\ne 0}_{\mu}(\bm{r})\rangle_{\overline{S}}=\int\frac{\mathrm{d}^3k}{(2\pi)^3}
    \Tr \tilde{G}^{a\ne 0}(\bm{k})=\frac{1}{4\pi}\int\frac{\mathrm{d}^3k}{(2\pi)^3}\frac{k^2+3\sigma m^2}{m^2(k^2+\sigma m^2)}.
\end{equation}

The first term in Eq.~(\ref{eq:20}) for general $N$ is given by 
\begin{align}
    \langle \cos\Tr (2\pi \Lambda^{-1}\Theta_\mu(\bm{r}) P)\rangle_{\overline{S}}&=\left\langle \cos\left[2\pi\Lambda^{-1}\frac{\theta_\mu^0(\bm{r})}{\sqrt{N}}+2\pi\Lambda^{-1}\sum^{N^2-1}_{a=1}\theta_\mu^a(\bm{r})\Tr (T^a P)\right]\right\rangle \\
    &=\exp\left[-\frac{2\pi^2\Lambda^{-2}}{N}\int\frac{\mathrm{d}^3k}{(2\pi)^3} \tilde{G}_{\mu\mu}^0(\bm{k})-2\pi^2\Lambda^{-2}\sum^{N^2-1}_{a=1}\Tr ^2(T^aP)\int\frac{\mathrm{d}^3k}{(2\pi)^3}\,\tilde{G}_{\mu\mu}^a(\bm{k})\right].
\end{align}
Using
\begin{align}
    \sum^{N^2-1}_{a=1}\Tr ^2(T^aP)=\sum^{N^2-1}_{a=1}T^a_{ij}P_{ji}T^a_{kl}P_{lk}=\left(\delta_{il}\delta_{jk}-\frac{1}{N}\delta_{ij}\delta_{kl}\right)P_{ji}P_{lk}=\Tr  P^2-\frac{1}{N}\Tr ^2P=1-\frac{1}{N},
\end{align}
we can express this by 
\begin{align}
     \langle \cos\Tr (2\pi\Lambda^{-1}\Theta_\mu(\bm{r}) P)\rangle_{\overline{S}}&=\exp\left\{-2\pi^2\Lambda^{-2}\int\frac{\mathrm{d}^3k}{(2\pi)^3}\,\left[\frac{1}{N}\tilde{G}^0_{\mu\mu}(\bm{k})+\left(1-\frac{1}{N}\right)\tilde{G}^{a\ne 0}_{\mu\mu}(\bm{k})\right]\right\}.
\end{align}
For a spherical integral region of $k$ (the region will be specified later), we can also rewrite this by
\begin{align}
  \langle \cos\Tr (2\pi\Lambda^{-1}\Theta_\mu(\bm{r}) P)\rangle_{\overline{S}}& =\exp\left\{-\frac{2\pi^2\Lambda^{-2}}{3}\int\frac{\mathrm{d}^3k}{(2\pi)^3}\,\left[\frac{1}{N}\Tr \tilde{G}^0(\bm{k})+\left(1-\frac{1}{N}\right)\Tr \tilde{G}^{a\ne 0}(\bm{k})\right]\right\}. \label{eq:21}
\end{align}
In the limit of $N\rightarrow 0$, the integrand of Eq.~(\ref{eq:21}) remains a finite function of $k$
\begin{align}
\lim_{N\to 0}\left[\frac{1}{N}\Tr \tilde{G}^0(\bm{k})+\left(1-\frac{1}{N}\right)\Tr \tilde{G}^{a\ne 0}(\bm{k})\right]=\frac{1}{4\pi}\frac{k^4+2(2\sigma+c)m^2k^2+3\sigma^2m^4}{m^2(k^2+\sigma m^2)^2}.
\end{align}

The second term of Eq.~(\ref{eq:20}) for general $N$ is given by
\begin{align}
   \frac{2\pi m^2}{V}\int \mathrm{d}^3r \sum_{a=0}^{N^2-1}\langle \left[\theta^a(\bm{r})\right]^2\rangle_{\overline{S}} =\frac{1}{2}\int\frac{\mathrm{d}^3k}{(2\pi)^3}\left[\frac{k^2+3(\sigma+Nc) m^2}{k^2+(\sigma+Nc) m^2}-\frac{k^2+3\sigma m^2}{k^2+\sigma m^2}\right]+\frac{N^2}{2}\int\frac{\mathrm{d}^3k}{(2\pi)^3}\frac{k^2+3\sigma m^2}{k^2+\sigma m^2}.
\end{align}
which vanishes in the replica limit. Thus, we obtain
\begin{equation}
    \lim_{N\to 0}\frac{2\pi m^2}{NV}\int \mathrm{d}^3r \sum_{a=0}^{N^2-1}\langle \left(\theta^a(\bm{r})\right)^2\rangle_{\overline{S}}=\int\frac{\mathrm{d}^3k}{(2\pi)^3}\,\frac{ cm^2 k^2}{(k^2+\sigma m^2)^2}.
\end{equation}

To summarize, we obtain $F(m)/V$ for  general $N$ as follows,
\begin{align}
    \frac{F(m)}{V}=&-\frac{6t\pi^{N-1}\Lambda^3}{\Gamma(N)}\exp\left\{-\frac{2\pi^2\Lambda^{-2}}{3}\int\frac{\mathrm{d}^3k}{(2\pi)^3}\,\left[\frac{1}{N}\Tr \tilde{G}^0(\bm{k})+\left(1-\frac{1}{N}\right)\Tr \tilde{G}^a(\bm{k})\right]\right\}\\
    &-\frac{1}{2}\int\frac{\mathrm{d}^3k}{(2\pi)^3}\left[\frac{k^2+3(\sigma+Nc) m^2}{k^2+(\sigma+Nc) m^2}-\frac{k^2+3\sigma m^2}{k^2+\sigma m^2}\right]-\frac{N^2}{2}\int\frac{\mathrm{d}^3k}{(2\pi)^3}\frac{k^2+3\sigma m^2}{k^2+\sigma m^2}\\
   &+\int\frac{\mathrm{d}^3 k}{(2\pi)^3}\,\ln\frac{k^2+(\sigma+Nc)m^2}{k^2+\sigma m^2}+N^2\int\frac{\mathrm{d}^3 k}{(2\pi)^3}\left[\ln m+\ln (k^2+\sigma m^2)\right],
\end{align}
where we use $\int \mathrm{d}|p(\bm{r})\rangle = \pi^{N-1}/\Gamma(N)$. This quantity vanishes linearly in $N$ in the replica limit with 
\begin{align}
    f(m) \equiv \lim_{N\to 0}\frac{F(m)}{NV}=&-\frac{6t\Lambda^3}{\pi}\exp\left[-\frac{\pi\Lambda^{-2}}{6}\int\frac{\mathrm{d}^3k}{(2\pi)^3}\,\frac{k^4+2(2\sigma+c)m^2k^2+3\sigma^2m^4}{m^2(k^2+\sigma m^2)^2}\right] \nonumber \\
    &-\int\frac{\mathrm{d}^3k}{(2\pi)^3}\,\frac{ cm^2 k^2}{(k^2+\sigma m^2)^2}+\int\frac{\mathrm{d}^3 k}{(2\pi)^3}\,\frac{cm^2}{k^2+\sigma m^2}.
\end{align}
Note that all three $k$-integrals on the right-hand side need the ultraviolet (UV) cutoff. Here, we use the following soft-cutoff function,
\begin{align}
    \Psi(\bm{k})=\frac{\Lambda^4}{(k^2+\Lambda^2)^2},
\end{align}
and replace the $k$-integrals by integrals with the cutoff function, i.e.
\begin{align}\label{eq:soft_cutoff}
    \int\frac{\mathrm{d}^3k}{(2\pi)^3}\,\cdots \to \int\frac{\mathrm{d}^3k}{(2\pi)^3}\,\Psi(\bm{k})\cdots.
\end{align}
The $k$-integrals are calculated as follows, 
\begin{align}
        &\int\frac{\mathrm{d}^3k}{(2\pi)^3}\,\frac{\Lambda^4}{(k^2+\Lambda^2)^2}\frac{k^4+2(2\sigma+c)m^2k^2+3\sigma^2m^4}{m^2(k^2+\sigma m^2)^2} \nonumber \\
        &=\frac{\Lambda^3}{8\pi}\frac{\Lambda^3+|m|\left[3\sqrt{\sigma}\Lambda^2+|m|\left(2c\Lambda+5\sigma\Lambda+3m\sigma^{3/2}\right)\right]}{m^2\left(\Lambda+\sqrt{\sigma}|m|\right)^3}, \end{align}
\begin{align}
    \int\frac{\mathrm{d}^3k}{(2\pi)^3}\,\frac{\Lambda^4}{(k^2+\Lambda^2)^2}\frac{ cm^2 k^2}{(k^2+\sigma m^2)^2}=\frac{1}{8\pi}\frac{cm^2 \Lambda^4}{\left(\Lambda+m\sqrt{\sigma}\right)^3},
\end{align}
\begin{align}
    \int\frac{\mathrm{d}^3 k}{(2\pi)^3}\,\frac{\Lambda^4}{(k^2+\Lambda^2)^2}\,\frac{cm^2}{k^2+\sigma m^2}=\frac{1}{8\pi}\frac{cm^2 \Lambda^3}{\left(\Lambda+m\sqrt{\sigma}\right)^2}.
\end{align}
Thus, we finally obtain,
\begin{equation}
    f=-\frac{6t\Lambda^3}{\pi}\exp\left\{-\frac{\Lambda}{48}\frac{\Lambda^3+|m|\left[3\sqrt{\sigma}\Lambda^2+|m|\left(2c\Lambda+5\sigma\Lambda+3m\sigma^{3/2}\right)\right]}{m^2\left(\Lambda+\sqrt{\sigma}|m|\right)^3}\right\}+\frac{1}{8\pi}\frac{c\sqrt{\sigma}m^3 \Lambda^3}{\left(\Lambda+m\sqrt{\sigma}\right)^3}.
\end{equation}

\clearpage
\twocolumngrid
\bibliographystyle{unsrt}
\bibliography{ref}

\end{document}